\documentclass [11pt]{article}

\usepackage{url}

\usepackage{draft} 
\usepackage{putex}
\usepackage{comment}

\usepackage{amsfonts}
\usepackage{graphicx}
\usepackage{amsmath,amssymb, dsfont}
\usepackage{latexsym}
\usepackage{mathrsfs}
\usepackage{bbold}
\usepackage{color}
\usepackage{slashed}
\usepackage{cancel}
\usepackage{soul} %
\usepackage{epsfig}
\usepackage{slashed}
\usepackage{epstopdf}
\usepackage{latexsym}
\usepackage{graphicx}
\usepackage{subcaption}
\usepackage{setspace} %
\usepackage{comment}
\usepackage[dvipsnames]{xcolor}
\usepackage{booktabs}
\usepackage{multirow}
\usepackage{ifthen,multicol,pifont,epsfig,latexsym,amsmath,amssymb,tikz-feynhand, tikz-feynman, feynmf,bbm}
\usepackage{cite}
\usepackage{caption}
\usepackage{slashed}
\usepackage{tabu}
\usepackage{xcolor}
\usepackage{epstopdf}
\usepackage{latexsym}
\usepackage{graphicx}
\usepackage{rotating}
\usepackage{booktabs}
\usepackage{bbm}
\usepackage{bbold}
\usepackage{enumitem}

\usepackage[numbers,compress]{natbib}

\definecolor{red}{rgb}{1,0,0}

\makeatletter
\def\section{\@startsection {section}{1}{\z@}{-3.5ex plus -1ex minus
 -.2ex}{2.3ex plus .2ex}{\large\bf}}
\def\subsection{\@startsection{subsection}{2}{\z@}{-3.25ex plus -1ex
minus -.2ex}{1.5ex plus .2ex}{\normalsize\bf}}
\makeatother
\makeatletter

\@addtoreset{equation}{section}

\makeatother

\def\be{\begin{equation}} \def\ee{\end{equation}}

\usepackage[colorlinks,linkcolor=black,citecolor=blue,urlcolor=blue,linktocpage,pagebackref]{hyperref}

\allowdisplaybreaks

\begin{document}

%\preprint{PUPT-2655}

\title{\centering Conformal QED in AdS as a BCFT}

\authors{Fabiana De Cesare \worksat{\INFN} and Simone Giombi \worksat{\PUJ}}

\institution{INFN}{INFN, Sezione di Torino, and Department of Physics, University of Turin, 10125, Turin, Italy}

\institution{PUJ}{Joseph Henry Laboratories, Princeton University, Princeton, NJ 08544, USA}

%\abstract{We study conformal Quantum Electrodynamics (QED) coupled to either $N_f$ massless fermions or $N_s$ conformally coupled scalars in Euclidean Anti-de Sitter (AdS$_d$) space for $d < 4$. Using the $\epsilon$-expansion around $d=4$, we investigate the associated Boundary Conformal Field Theories (BCFTs) defined by imposing either Dirichlet or Neumann boundary conditions on the gauge field. We compute the regularized AdS free energy at the interacting fixed point up to next-to-leading order, finding that while the Neumann free energy is larger than the Dirichlet free energy close to $d=4$, this ordering inverts at a critical dimension $3 < d_\mathrm{crit} < 4$. Furthermore, we extract some boundary conformal data at one loop, including the anomalous dimensions of the lightest singlet scalar operators. As a consistency check, we explicitly recover the protected dimension of the displacement operator. More importantly, our results indicate that while the Neumann boundary condition defines a stable BCFT that safely interpolates down to $d=3$, the Dirichlet boundary condition becomes unstable and annihilates with another boundary fixed point at a critical dimension $3 < d_\mathrm{merg} < 4$. }
%Our one-loop results suggest that $d_\mathrm{merg} < d_\mathrm{crit}$ in all cases considered.

\abstract{We study conformal Quantum Electrodynamics (QED) coupled to either $N_f$ massless fermions or $N_s$ conformally coupled scalars in Euclidean Anti-de Sitter (AdS$_d$) space for $d < 4$. Using the $\epsilon$-expansion around $d=4$, we investigate the associated Boundary Conformal Field Theories (BCFTs) defined by imposing either Dirichlet or Neumann boundary conditions on the gauge field. We compute the regularized AdS free energy at the interacting fixed point up to next-to-leading order and extract some of the boundary conformal data at one loop, including the anomalous dimensions of the lightest singlet scalar operators. For Dirichlet boundary conditions, extrapolation of our $\epsilon$-expansion results indicate that one of these scalar operators, which is irrelevant near $d=4$, reaches marginality within the range $3 < d < 4$. This suggests that the Dirichlet boundary condition may not define a stable BCFT in $d=3$. 
%Our one-loop results suggest that $d_\mathrm{merg} < d_\mathrm{crit}$ in all cases considered.
}

\date{}

\maketitle

\tableofcontents

\section{Introduction}
Boundary conformal field theories (BCFTs) are a natural extension of conformal field theories (CFTs) that describe critical phenomena in the presence of boundaries \cite{Diehl:1996kd,McAvity:1995zd,Liendo:2012wv}. A powerful approach to studying BCFTs involves placing the corresponding bulk CFT in Anti-de Sitter (AdS) space. Because AdS is Weyl equivalent to the flat half-space, BCFT correlation functions can be mapped directly to those in AdS via a simple Weyl rescaling of the operators. This allows to apply the technical framework developed in the AdS/CFT literature to extract new results about BCFTs. This strategy has been successfully employed to compute boundary conformal data in the $\epsilon$ and large $N$ expansions for various interacting theories, including the $O(N)$ Wilson-Fisher and Gross-Neveu universality classes \cite{Carmi:2018qzm,Giombi:2020rmc,Giombi:2021cnr,Giombi:2025pxx,Diatlyk:2026eta,Csipes:2026nyo}. Remarkably, some of these perturbative computations have been successfully matched with non-perturbative results obtained from Monte-Carlo simulations and fuzzy sphere methods,  showing the power of the AdS approach.

Defining a quantum field theory in AdS requires the specification of boundary conditions (BCs). Different choices of conformal BCs define distinct BCFT universality classes. By perturbing the action with relevant boundary operators, one can trigger boundary Renormalization Group (RG) flows that connect these fixed points. A fundamental property expected to govern such trajectories is the boundary $F$-theorem, proposed in \cite{Giombi:2020rmc} following analogous conjectures for bulk CFTs \cite{Giombi:2014xxa} and defect CFTs \cite{Kobayashi:2018lil}. This theorem states that the properly regularized AdS free energy, defined as
\begin{equation}
    \tilde{F} = -\sin\left(\frac{\pi(d-1)}{2}\right) F_{\text{AdS}_d} \,,
    \label{eq:bF_theorem}
\end{equation}
 must decrease along a boundary RG flow, ensuring that $\tilde{F}_{\text{UV}} > \tilde{F}_{\text{IR}}$. We note that an analogous inequality can be derived even when the bulk theory is not conformal, as shown by \cite{Bason:2025sxb}, though this will not enter in this work as we will restrict our attention entirely to the case of conformality in the bulk. A closely related quantity \cite{Kobayashi:2018lil} $\tilde{s}=-\sin\left(\frac{\pi(d-1)}{2}\right) \left(F_{\text{AdS}_d}-\frac{1}{2}F_{S^d}\right)$, which smoothly interpolates between boundary central charges and $g$-functions and unifies the corresponding monotonicity theorems, was used in \cite{Giombi:2025pxx, Diatlyk:2026eta} to obtain estimates for boundary central charges in the 3d $O(N)$ and Gross-Neveu-Yukawa models. In this work we will be interested in comparing the values of $\tilde{F}$ between boundary conditions, and hence the difference between $\tilde{F}$ and $\tilde{s}$ will not play an important role. 
 
A universal feature of any BCFT is the breaking of translational invariance in the direction orthogonal to the boundary. This broken symmetry implies the existence of a distinguished scalar operator localized on the boundary, known as the displacement operator, which appears as a contact term in the Ward identity associated with these broken translations. Because of its relationship to the conserved bulk stress tensor, which has a protected dimension equal to $d$ when the bulk is conformal, the dimension of the displacement operator is also protected and strictly equal to $\Delta = d$. The presence of a boundary scalar with this exact dimension is thus a defining signature of a BCFT.
%Because of its  relationship to the conserved bulk stress tensor, its conformal dimension is protected and strictly equal to the bulk spacetime dimension, $\Delta=d$. The presence of this operator is  a defining signature of conformality in the bulk.

In this paper, we consider conformal Quantum Electrodynamics (QED) %coupled to massless fermions and conformally coupled scalars 
in AdS$_d$ for $d<4$. We use the $\epsilon$-expansion around $d=4$ to compute perturbative observables at the interacting infrared fixed point under different choices of BCs for the gauge fields. 
 Beyond its formal interest, the study of conformal boundary conditions in QED is strongly motivated by its applications in condensed matter physics. In three dimensions, conformal QED emerges as the low-energy description of strongly correlated systems, particularly in the context of the Deconfined Quantum Critical Point (DQCP) \cite{Senthil:2003eu, Senthil:2004qcv} and Dirac spin liquids \cite{Marston:1989, Franz:2002, Herbut:2002, Hermele:2005ks, Song:2018ccq, Song:2019tbj}. Since experimental realizations of these systems inherently possess physical surfaces, understanding the possible conformal boundary conditions is a necessary step to theoretically capture their finite-size effects and surface critical phenomena.

To see how different choices of boundary conditions arise, we first note that a free massless vector field $A_\mu$ propagating in Euclidean AdS$_d$ generically admits two possible boundary conditions when $3<d\le4$. If we use Poincaré coordinates with metric
%To see how these choices arise, we first note that a free massless vector field $A_\mu$ propagating in Euclidean AdS$_d$ generically admits two possible boundary conditions when $3<d\le4$. If we use Poincaré coordinates with metric
\begin{equation}
d s^2=\frac{d z^2+d\vec{x}^2}{z^2}\,, 
\end{equation}
we have either Neumann (N) BC, for which the bulk gauge field asymptotes to a dynamical gauge field at the boundary,
\begin{equation}
    A_i(x)\underset{z\to0}{\sim} a_i(\vec{x})\,,
\end{equation}
or  Dirichlet  (D) BC, for which the bulk field is dual to a conserved current at the boundary, 
\begin{equation}
    A_i(x)\underset{z\to0}{\sim} z^{d-3}\, e^2 J_i(\vec{x})\,.
\end{equation}
 As usual for gauge theories, we need to impose a gauge fixing and introduce ghosts $c$, which also need the boundary condition to be specified. The two possible conditions $c\sim z^{\delta_\pm}$, with $\delta_-=0$ and $\delta_+=d-1$, are constrained to the choice of boundary condition on the physical vector fields: with Neumann BC, the presence of dynamical gauge fields at the boundary requires the gauge transformation (and equivalently the ghost field) to persist at the boundary ($c\sim z^{0}$); with Dirichlet BC, the gauge transformation should instead decay faster than the gauge field at the boundary, as the bulk gauge field is dual to a gauge-invariant current in this case ($c\sim z^{d-1}$).
 
The dynamics of gauge theories in an AdS background have been recently explored in the context of gauge theories in four dimensions, in the absence of conformality in the bulk \cite{Aharony:2012jf,Ankur:2023lum,Copetti:2023sya,Ciccone:2024guw,Ciccone:2025dqx,DiPietro:2025ozw,Ankur:2026ylr}. In this setup, the study of conformal data associated to different BCs provides important insights into the mechanisms of confinement, chiral symmetry breaking  and the ending of the conformal window in flat space.  A fundamental concept arising from these studies is that, in $d=4$, the Dirichlet BC is expected to become unstable and disappear because it would otherwise give rise to colored asymptotic states in flat space. The most compelling scenario to explain this disappearance is that the Dirichlet BC merges and annihilates with another BC, denoted in the literature as D$^*$ \cite{Ciccone:2024guw,Ciccone:2025dqx}. 
Specifically, this annihilation is triggered by an irrelevant scalar operator at the boundary becoming marginal at a critical AdS radius, or equivalently, at a critical value of the bulk coupling $g^2_{\text{crit}}$. As explained in \cite{Lauria:2023uca,Ciccone:2024guw}, the beta function for the boundary coupling $\eta$ associated with that operator takes the form
\begin{equation}
\beta_\eta = c_1\eta^2 + c_2\left(\frac{1}{g^2} - \frac{1}{g^2_{\text{crit}}}\right)\,,\quad  g^2 \lesssim g^2_{\text{crit}}\,,
\label{eq:beta_eta}
\end{equation}
where the coefficients $c_{1,2}$ are determined by the data of the boundary CFT, which are unknown if $g^2_{\text{crit}}$ is not small. However, the relevant information lies in their relative sign, as it determines whether the two fixed points corresponding to the zeros of this beta function are real or complex. Since the Dirichlet BC is known to exist for $g^2 < g^2_{\text{crit}}$, the D$^*$ boundary condition must also be present in this regime. Conversely, for values of $g^2 > g^2_{\text{crit}}$, the fixed points become complex, providing a natural explanation for why the D boundary condition ceases to be a viable conformal BC.

 On the other hand, the Neumann BC, which corresponds to a confining theory even at a small radius, can smoothly approach the flat-space limit. An interesting direction for testing this scenario involves placing a Wilson line at the AdS boundary with Neumann BC and comparing its behavior at small radius, where it can be studied as a weakly coupled conformal defect, with its behavior at large radius, where it corresponds to a flat-space confining string. Thus far, this framework has been explored in $d=3$ \cite{Gabai:2025hwf,Gabai:2026myo}, yielding promising results which suggest a smooth interpolation between the two regimes. 
 %Within this setup, one could also extract the masses of the lightest glueballs and compare them with lattice results by tracking the asymptotics of the scaling dimensions of scalar singelt boundary operators. 
 %A possibly effective tool for evaluating the continuous evolution of boundary data from weak to strong coupling is a recently developed framework that formulates the flow as a system of ordinary differential equations \cite{Loparco:2026fki,DeCesare:2026bor}.
 
  In this work, we investigate what is the fate of D and N boundary conditions in the abelian case for $d < 4$ when the bulk reaches conformality. To obtain bulk conformality, the gauge fields must be coupled to matter. One possibility is to introduce massless fermions and study the corresponding infrared fixed point in $d=4-\epsilon$ using the $\epsilon$-expansion, see, for example, \cite{DiPietro:2015taa, Giombi:2015haa} for extensive studies of the bulk CFT. Alternatively, one can couple the gauge field to a sufficiently large number of scalars \cite{Halperin:1973jh, Moshe:2003cg}. In an AdS background, these cannot simply be minimally coupled massless scalars, as this would break conformal invariance at zero coupling, and must be taken to be conformally coupled instead. A scalar quartic self-interaction is also necessary to reach an IR fixed point of the bulk CFT.  
  
For both the fermionic and scalar cases, we compute the regularized AdS free energy at the interacting fixed point up to next-to-leading order. Our results indicate that, while the Neumann free energy is larger than the Dirichlet free energy close to $d=4$, this ordering inverts at a critical dimension $3 < d_\mathrm{crit} < 4$.
  
Furthermore, we extract some boundary conformal data at one loop, including the anomalous dimensions of the two lightest singlet scalar operators at the boundary. As a consistency check, we explicitly recover that one of these is the displacement operator. More importantly, we find that the other operator has positive anomalous dimension for Neumann BC and instead negative anomalous dimension for Dirichlet BC, reaching marginality at a critical dimension $3 < d_\mathrm{merg} < 4$. This suggests that while the Neumann BC defines a stable BCFT that safely interpolates down to $d=3$, the Dirichlet BC becomes unstable and may annihilate with another boundary fixed point at $d=d_\mathrm{merg}$. To understand why reaching marginality is expected to trigger merger and annihilation, the argument described in \cite{Lauria:2023uca} cannot be used directly: their mechanism relies on a running bulk coupling, while we evaluate the boundary data exactly at the bulk fixed point $e=e^*$ as a function of $\epsilon$. However, for any fixed $\epsilon$, this IR BCFT is reached by flowing from the free bulk theory in the UV. Suppose, for the sake of contradiction, that for some $\epsilon$ the flow successfully reached a Dirichlet BCFT where this singlet scalar operator is below marginality. Since this operator is irrelevant in the UV, continuity dictates that it must cross marginality at some finite AdS radius along the flow, or equivalently, at a critical intermediate coupling $e_{\text{crit}} < e^*$. At this point, the system would encounter the exact same instability described previously for the non-abelian theory in $d=4$, see Eq. \eqref{eq:beta_eta}. This instability would drive a merger and annihilation, preventing the theory from ever reaching the putative Dirichlet fixed point. This provides an indication that the Dirichlet BCFT ceases to exist when the operator crosses marginality.  
  
Interestingly, our one-loop results suggest that $d_\mathrm{merg} < d_\mathrm{crit}$ in all cases considered. Assuming the Neumann boundary condition represents the stable IR fixed point reached after the merger, this is consistent with the boundary $F$-theorem in Eq. \eqref{eq:bF_theorem}. Indeed, in order to have a flow from the merging BCs to the Neumann BC at $d=d_\mathrm{merg}$, we need the relations $\tilde{F}_{\mathrm{D}^*} = \tilde{F}_\mathrm{D}$ and $\tilde{F}_{\mathrm{D}^*} > \tilde{F}_\mathrm{N}$ to be satisfied at this point. This naturally implies $d_\mathrm{merg} < d_\mathrm{crit}$.
  
The structure of this paper is as follows. In Section \ref{sec:free_energy}, we evaluate the leading-order free energy of abelian gauge theories coupled to $N_f$ Dirac fermions in AdS$_d$ for both Dirichlet and Neumann boundary conditions for the gauge field. In Section \ref{sec:free_energy_NL}, we develop the AdS Feynman rules for fermionic QED and compute the next-to-leading order corrections to the free energy at the interacting fixed point. Section \ref{sec:BCFTferm} is devoted to extracting the anomalous dimensions of the lightest singlet scalar operators. In Section \ref{sec:EOMferm}, we compute the anomalous dimensions of the boundary fermion with Dirichlet BC for the gauge field using the bulk equations of motion. Finally, in Section \ref{sec:scalar_QED}, we extend our analysis to the case of QED coupled to $N_s$ conformally coupled scalars. 
 
 %Let us consider the case in which the bulk gauge theory is coupled to $N$ matter fields, with $N$ sufficiently large to ensure conformality in the bulk. Neumann and Dirichlet BC correspond then to two different CFTs, which, at least at large $N$ \FDC{What about finite $N$?}, are respectively the UV and the IR endpoint of an RG flow generated by the double-trace operator $J_\mu J^\mu$, with $J_\mu$ having dimension $\Delta$. This operator is irrelevant for $\Delta=\Delta_+$ and becomes marginal at $\Delta=(d-1)/2$. 

\section{Free energy at the leading order}
\label{sec:free_energy}

Let us consider an abelian gauge theory with $N_f$ massless charge-one Dirac fermions %in the fundamental representation 
on AdS$_d$ with Dirichlet and Neumann BC for the gauge field.
We want to compute the free energy in $d=4-\epsilon$, defined as
\be
F  = -\log  Z_{\text{AdS}_d}~, \qquad 
Z_{\text{AdS}_d}  = \frac{1}{\mathrm{vol}(\mathcal{G})} \int \mathcal{D}A\mathcal{D}\psi\mathcal{D}\bar{\psi} \, \exp\left(-S(A,\psi,\bar{\psi}, g)\right)~.
\label{eq:a}
\ee
Here $g$ denotes the round metric $g_{\mu\nu}$ on AdS$_d$ with radius $1$ and coordinate $x$, while $\mathrm{vol}(\mathcal{G})$ is the volume of the group of gauge transformations. We can split the action in
\begin{align}
\begin{split}
S=S_\text{Maxwell}+S_\text{Ferm}+S_\text{curv}\ ,
\label{eq:b}
\end{split}
\end{align}
with
\begin{align}
S_\text{Maxwell} & =\int d^{d} x \sqrt{g}\left(\frac{1}{4 e_{0}^{2}}  F_{\mu\nu}(x)F^{\mu\nu}(x)\right)\ \label{eq:c1} \,, \\
S_\text{Ferm} &=\int d^{d} x \sqrt{g}\left(-\sum_{i=1}^{N_f} \bar{\psi}_{i} \gamma^{\mu}\left(\nabla_{\mu}+i A_{\mu}\right) \psi^{i}\right)\ ,\label{eq:c2} \\
S_\text{curv} & =\int d^{d} x \sqrt{g}\left(b_{0} E +c_{0} \mathcal{R}^{2} /(d-1)^{2}\right),
\label{eq:c3}
\end{align}
where $e_0$ is the bare electric coupling constant, $\psi^{i}$ are $N_f$ four-component Dirac fermions\footnote{We dimensionally continue the theory while keeping the number of fermion components fixed, as it was done on the sphere in Ref.\cite{Giombi:2015haa}.} and $\nabla_{\mu}$ is the curved space covariant derivative which includes the spin connection term when acting on fermions.
  As the action should contain all operators that are marginal in $d=4$, we have added the curvature terms together with their bare coupling parameters $b_{0}$ and $c_{0}$.\footnote{In a generic Euclidean manifold we should also include a term with the square of the Weyl tensor, omitted here as it vanishes.} For future purposes, we recall the expression for the Ricci scalar $\mathcal{R}$ and the Euler density $E$ on AdS$_d$:
\begin{equation}
\begin{aligned}
&\mathcal{R}=-{d(d-1)}{}~,~~E=\mathcal{R}_{\mu \nu u \rho} \mathcal{R}^{\mu \nu u \rho}-4 \mathcal{R}_{\mu \nu} \mathcal{R}^{\mu \nu}+\mathcal{R}^{2}={d(d-1)(d-2)(d-3)}~.
\label{eq:curvScala}
\end{aligned}
\end{equation}

\subsection{One-loop determinants}
\label{sec:oneloop}

At leading order in the loop expansion the free energy is determined by one-loop determinants. 
As a consequence of the splitting in eq.~\eqref{eq:b}, we can divide the leading term of the free 
energy $F_\text{Free}$ in three parts:
\begin{equation}
F_\text{Free}=F_\text{Maxwell}+F_\text{free-ferm}+F_\text{curv}\ ,
\end{equation}
with
\begin{align}
F_\text{Maxwell} & =-\log\bigg(\frac{1}{\mathrm{vol}(\mathcal{G})} \int \mathcal{D}Ae^{-S_{\text{Maxwell}}[A, h]}\bigg)
\label{eq:d1}\,, \\
F_\text{free-ferm} & =-\log \Big(\int\mathcal{D}\psi\mathcal{D}\bar{\psi} \, e^{-S_{\text{free-ferm}}[\psi, h]}\Big)\,,
 \label{eq:d2} \\
F_\text{curv} & =\text{Vol}(H^d) R^{d-4}(d(d-1)(d-2)(d-3))b_0+ d^2 c_0),
\label{eq:d3}
\end{align}
where $S_{\text{free-ferm}}$ is the free fermion action and 
$\text{Vol}(H^d)=\pi^{d-1}\Gamma(\frac{d-1}{2})$ is the regularized volume of the $d$-dimensional hyperbolic space with unit radius.

\paragraph{Dirichlet BC} We start from the computation of $F_\text{Maxwell}$ with Dirichlet boundary conditions for the gauge field. Following the analogous computation performed in \cite{Giombi:2015haa}, we can express this free energy in terms of one-loop determinants as 
%\FDC{I am not so sure about the coupling normalization... the $\log e^2$ should cancel in some way}
\begin{equation}\label{eq:fmax}
F_\text{Maxwell}=\frac{1}{2}\log\det{}_{(1)}\left({-\nabla^2-(d-1) }\right)-\frac{1}{2}\log\det {}_{(0)}\left({-\nabla^2}\right),
\end{equation}
%\be
%F_\text{Maxwell}=\frac{1}{2}\log\det{}_{(1)}\left(\frac{-\nabla^2-(d-1) }{2 G e^2}\right)-\frac{1}{2}\log\det {}_{(0)}\left({-\nabla^2}\right),
%\ee
where the subscripts $(1)$ and $(0)$ indicate that the determinant is taken on the space of transverse vector fields and on scalar fields respectively, the latter corresponding to the ghost contribution. Note that, contrary to what happens on the sphere, here the free energy does not include a term $\sim\log e^2$ because gauge theories with Dirichlet BC do not have ghost zero modes. The one-loop determinants in Eq.~\eqref{eq:fmax} were derived for all integer spins and all dimensions by Camporesi and Higuchi \cite{Camporesi:1993mz,Camporesi:1994ga} and read
\begin{equation}\label{eq:zeta}
\begin{aligned}
&F^{(s)}(\Delta)= \log \det {}_{(s)}\left(-\nabla^2+\Delta(\Delta-(d-1))-s\right)\\
& F^{(s)}(\Delta)= \frac{\operatorname{Vol}_{H^{d}}}{\operatorname{Vol}_{S^{d-1}}} 2^{d-2} g(s) \int_0^{\infty} d u \ {\mu_s(u)}\log{\left(u^2+\left(\Delta-\frac{d-1}{2}\right)^2\right)}\,,
\end{aligned}
\end{equation}
where $g(s)$ is the number of degrees of freedom of a symmetric traceless transverse spin $s$ field in $d$ dimensions and $\mu_s(u)$ is the so-called spectral density:
\begin{equation}
\begin{aligned}
   & g(s)=(2 s + d - 3) \frac{\Gamma(s + d - 3)}{\Gamma(d - 2) \Gamma(s + 1)}\,,\\
   &\mu_s(u)= \frac{\left(u^2 + (s + \frac{d - 3}{2})^2\right)}{\left(2^{d - 2} \Gamma\left(\frac{d}{2}\right)\right)^2 }\frac{\Gamma\left(
  iu + \frac{d - 3}{2}\right)\Gamma\left(-iu + \frac{d - 3}{2}\right)}{\Gamma(iu) 
 \Gamma(-iu)}\,.
\end{aligned}
\end{equation}
Going back to Eq.~\eqref{eq:fmax}, we then get
\begin{equation}\label{eq:Maxwell2}
F_\text{Maxwell}=\frac{1}{2}F^{(1)}(d - 2)-\frac{1}{2}F^{(0)}(d - 1)\,.
\end{equation}
We now set $d=4-\epsilon$ and compute the $\epsilon$-expansion of this quantity. We begin by noting that the integrals over $u$ defining $F^{(s)}(\Delta)$ are divergent at large values of $u$. To handle these divergences, we generalize the procedure described in App.~A of Ref.~\cite{Giombi:2025pxx}, applied in that case to the free energy of conformally coupled scalars in AdS.

Let us consider the function $\mu_s(u)$ multiplying the logarithmic factor $\log(u^2 + M_s^2)$ in \eqref{eq:zeta}, where the effective mass squared is $M_1^2 = {(d-3)^2}/{4}$ for the transverse vector and $M_0^2 = {(d-1)^2}/{4}$ for the ghost. To isolate the divergence, we replace $\mu_s(u)$ with its large-$u$ asymptotic expansion, $\hat{\mu}_s(u)$, constructed as a series in powers of $(u^2 + M_s^2)$. We then rewrite the integral by adding and subtracting this asymptotic expansion. The subtracted integral, containing the remainder $\mu_s(u) - \hat{\mu}_s(u)$, is finite in $d=4$, thus we can safely expand its integrand in powers of $\epsilon$ and evaluate it numerically. The  divergence is now entirely isolated within the integral of the asymptotic piece $\hat{\mu}_s(u)$, which can be computed analytically by means of dimensional regularization with the integral
\begin{equation}
    \int_0^\infty d u \ (u^2 + M^2)^a \log(u^2 + M^2) = \frac{d}{d a} \left( \frac{\sqrt{\pi} M^{2a+1} \Gamma\left(-a-\frac{1}{2}\right)}{2 \Gamma(-a)} \right)\,.
\end{equation}
Evaluating this analytic result for the individual terms in $\hat{\mu}_s(u)$, expanding in powers of $\epsilon$, and combining it with the numerical evaluation of the finite remainders, we obtain the following results for the transverse vector and the ghost:
\begin{equation}
\begin{aligned}
    F^{(1)}(d-2) &= \frac{11}{30\epsilon} - 0.03842 + 0.55140\epsilon + \mathcal{O}(\epsilon^2)\,, \\
    F^{(0)}(d-1) &= -\frac{29}{90\epsilon} - 0.01158 - 0.53026\epsilon + \mathcal{O}(\epsilon^2)\,.
\end{aligned}
\end{equation}
Replacing these in \eqref{eq:Maxwell2}, we arrive at the final Maxwell free energy with Dirichlet boundary conditions:
\begin{equation}\label{eq:maxD}
   { F^\mathrm{D}_\mathrm{Maxwell}}= \frac{31}{90 \epsilon} -0.013417+
0.54083\epsilon\,+\mathcal{O}\left(\epsilon^2\right),
\end{equation}
 We note that the coefficient of the $1/\epsilon$ term represents the contribution of the conformal anomaly integrated on the AdS background. As a check, we can compare it with the corresponding anomaly pole on the sphere \cite{Giombi:2015haa}. Since the integrated anomaly on AdS is half that of the sphere due to the ratio of their regularized volumes, the quantity $F_{\mathrm{AdS}_d}-1/2 F_{S_d}$ must be finite \cite{Giombi:2025pxx,Bason:2025sxb}, which is perfectly consistent with our result. 
%We note that the coefficient of the $1/\epsilon$ term coincides with the conformal anomaly of a vector field in AdS. As a check, we can compare it with the same quantity on the sphere, see \cite{Giombi:2015haa}. As explained for instance in \cite{Bason:2025sxb}, the quantity $F_{\mathrm{AdS}_d}-1/2 F_{S_d}$ must be finite, which is indeed consistent with our result. 

%This results into
%\begin{equation}\label{eq:maxD}
 %  { F^\mathrm{D}_\mathrm{Maxwell}}= \frac{31}{90 \epsilon} -0.013417+0.54083\epsilon\,+\mathcal{O}\left(\epsilon^2\right),\end{equation}which is again consistent with the conformal anomaly for a vector field \cite{Giombi:2015haa} and with the fact that $F_{\mathrm{AdS}_d}-1/2 F_{S_d}$ must be finite.
%\FDC{With respect to eq. 2.4 of ref. \cite{Giombi:2013fka}, in eq.\eqref{eq:zeta} we omitted the logarithmic term $\zeta_{(\Delta, s)}(0)\log(\Lambda^2)$, which is related to the conformal anomaly. In dimensional regularization this term should arise as a pole in $\epsilon$ from the computation of $\zeta_{(\Delta, s)}^{\prime}(0)$. }

\paragraph{Neumann BC} Computing the value of the free energy with Neumann boundary condition is more nontrivial. We note that in this case the boundary condition for the ghosts allows for zero modes, which then have to be removed, as it happens on the sphere. This makes it harder to evaluate the difference in the free energy between Dirichlet and Neumann boundary conditions, which we denote by
\begin{equation}
    \delta F_\mathrm{Maxwell}
    ={ F^\mathrm{N}_\mathrm{Maxwell}}-{ F^\mathrm{D}_\mathrm{Maxwell}}\,.
\end{equation}
For massive vector fields, the same quantity can be evaluated by first  differentiating the difference with respect to the mass, then using zeta function regularization and finally integrating back the result, see \cite{Giombi:2013yva}. However, the presence of ghost zero modes in the massless case leads to a divergence in the last integral which makes it difficult to follow this procedure consistently. 

An alternative indirect way to proceed is to use known results for the change in the free energy due to double trace deformations. To see how this works, we draw an analogy from scalar fields in AdS. In this case, one can obtain Neumann BC from Dirichlet BC by adding the deformation $\int_{\partial AdS}d^{d-1}x \ \sigma O$ to the boundary, where $O$ is the dual operator to the Dirichlet bulk field,  and $\sigma$ is a boundary auxiliary field which we promote to be dynamical. The difference between the Neumann and Dirichlet free energy can then be expressed as $F^N_{s}- F_{s}^D = F_{\sigma}$, where 
\begin{equation}
    F_{\sigma} = \frac{1}{2} \log \det \langle OO \rangle
\end{equation} 
is the sphere free energy of a boundary field with induced kinetic term $\langle OO \rangle$. 

This logic naturally extends to the Maxwell case. First, consider the theory with Dirichlet boundary conditions and take the generating functional for the correlation functions of the boundary conserved current $J_i$. This will be a function of the non-dynamical boundary value of the gauge field, $A|_{\partial AdS} = a_i$, as in
\begin{equation}
Z_D[a_i] = \langle e^{\int_{\partial AdS} d^{d-1}x \, a_i J_i} \rangle_D\,.
\end{equation}
The partition function of the theory with Neumann boundary conditions can then be obtained by promoting the boundary gauge field $a_i$ to be dynamical \cite{Witten:2003ya}, leading to the relation:
\begin{equation}
Z_N = \int \frac{\mathcal{D}a_i}{\operatorname{vol}({G})} Z_D[a_i]\,,
\end{equation}
where $\operatorname{vol}({G})$ is the volume of boundary gauge transformations.
To evaluate $Z_N$, we simply expand $Z_D[a_i]$ in powers of $a_i$\begin{equation}
Z_D[a_i] = Z_D[0] \, e^{-\frac{1}{2} \int a_i K_{ij} a_j + \mathcal{O}(a_i^3)}\,,
\end{equation}
where $Z_D[0]$ is the pure Dirichlet partition function (with $a_i=0$), and the kernel $K_{ij}$ is determined by the boundary two-point function $K_{ij} = -\langle J_i J_j \rangle_D$. Substituting this expansion back into the path integral gives the ratio:
\begin{equation}
\frac{Z_N}{Z_D[0]} = \int \frac{\mathcal{D}a_i}{\operatorname{vol}(\mathcal{G})} e^{-\frac{1}{2} \int a_i K_{ij} a_j + \mathcal{O}(a_i^3)}\,.
\end{equation}
Taking the logarithm, we obtain the free energy difference as
%$$\delta F_{\text{Maxwell}} = F^N_{\text{Maxwell}} - F^D_{\text{Maxwell}} = F_{a_i}\,,$$where $F_{a_i}$ is the free energy (the path integral over the boundary field) of the boundary gauge field on $S^{d-1}$ governed by the induced quadratic term $K_{\mu\nu}=-\langle J_\mu J_\nu\rangle$.
%This logic naturally extends to the Maxwell case: going from Dirichlet to Neumann boundary conditions corresponds to coupling the boundary gauge field $a_i$ to the conserved current $J_i$ via $\int_{\partial AdS}d^{d-1}x\  J_i a_i$ and promoting $a_i$ to a dynamical field \cite{Witten:2003ya}. Consequently, the free energy difference is given by 
\begin{equation}
    \delta F_{\text{Maxwell}} = F_{a_i}\,,
\end{equation} 
where $F_{a_i}$ is the free energy of the boundary gauge field on $S^{d-1}$ with a quadratic term governed by $K_{ij}$. This is essentially the calculation performed in Sec.3 of \cite{Giombi:2015haa}, with a slightly different interpretation. In our setup, we are evaluating this on $S^{d-1}$ instead of $S^d$, and the normalization $C_J$ appearing in their Eq. (3.5) is fixed in terms of the bulk coupling constant $e_0$, by imposing the  conserved current Ward identity. This can be obtained using standard AdS/CFT prescriptions and reads \cite{Freedman:1998tz}\footnote{To correctly match the result of \cite{Freedman:1998tz}  with the definition of $C_J$ in Eq.(3.5) of    \cite{Giombi:2015haa}, one should take $B$ below Eq.(54) of \cite{Freedman:1998tz},  multiply it by $2(d-2)(d-1)/(2\pi)^d$, and finally replace $d$ with $d-1$.}
\begin{equation}
   C_J= \frac{2^{d-3}(d-3)  \Gamma\left(\frac{d}{2}\right)}{e_0^{2}\pi^\frac{d}{2} }\,.
\end{equation}
We thus have
\begin{equation}
F_{a_i}=\frac{1}{2} \log \operatorname{det}_T\left(\frac{K_{ij}}{2 \pi}\right)-\frac{1}{2} \log \operatorname{det}^{\prime}\left(-\nabla^2\right)+\log \left(2 \pi \sqrt{\operatorname{vol}\left(S^{d-1}\right)}\right)\,,
\end{equation}
where the subscript ‘T’ indicates that the determinant is taken on the space of transverse vector
fieds and the prime means zero mode excluded. The last two terms arise from the gauge fixing at the boundary and are  obtained by following the procedure described in Sec. 2 of \cite{Klebanov:2011td}.  Adapting the results of \cite{Giombi:2015haa} to our specific case, we get 
\begin{equation}
F_{a_i}=\frac{1}{2} \log \left( \frac{8\pi^{d} \, C_J \, \Gamma\left({4-d}\right)\sin\left(\frac{\pi d}{2 }\right)}{ \, {2-d} } \right)+\int_0^1 du \ f_d(u)\,,
%F_{a_i}=\frac{1}{2} \log \left( \frac{4(d-2)\pi^{\frac{d+2}{2}} \, 2^{5-d} \, \Gamma\left(\frac{5-d}{2}\right) \Gamma(d-2)}{e_0^2 \, \Gamma\left(\frac{d-3}{2}\right) \Gamma\left(\frac{d}{2}\right)} \right)+\int_0^1 du f(u,d)\,,
\end{equation}
with
\begin{equation}
\begin{aligned}
f_d(u) = \frac{1}{2} \Bigg( \frac{1}{u-1} - \frac{1}{u} \Bigg) + \frac{\sec\left(\frac{d\pi}{2}\right)}{16 \Gamma(d)} \Bigg(8(d-2u-1) \cos\left(\frac{\pi(d-2u)}{2}\right) \Gamma(d-u-1) \Gamma(u) &\\
 - (d-3)^2(d-2)u \Big( (d-3)^2 u^2 - d^2 + 2d - 5 \Big) \Gamma\left(\frac{(d-3)(1-u)}{2}\right) &\\ \times\left(\frac{(d-3)(u+1)}{2}\right) \sin\left(\frac{(d-3)\pi u}{2}\right)\Bigg)\,.&
\end{aligned}
\end{equation}
Expanding for $d=4-\epsilon$ and integrating over $u$ term by term, we get
\begin{equation}
    \delta F_\mathrm{Maxwell}=\frac{1}{2}\log(2\pi) - 2.405 \epsilon -\frac{1}{2} \log e_0^2\,+\mathcal{O}\left(\epsilon^2\right),
\end{equation}
where the term proportional to $\log e_0^2$ is consistent with the presence of zero modes for the ghost with this choice of boundary condition.

\paragraph{Fermions} The expression for $F_{\text{free-ferm}}$ was found in ref.~\cite{Giombi:2021cnr}. The result for a single Dirac spinor is 
\begin{equation}
F_\mathrm{free-ferm}=-\frac{\operatorname{Vol}\left(H^d\right) c_d}{(4 \pi)^{\frac{d}{2}} \Gamma\left(\frac{d}{2}\right)} \int_0^{\infty} d u \frac{\left|\Gamma\left(\frac{d}{2}+i u\right)\right|^2 \log \left( u^2\right)}{\left|\Gamma\left(\frac{1}{2}+i u\right)\right|^2}\,,
\label{eq:eqferm}
\end{equation}
where $c_d$ is the number of components of the Dirac spinor, which we here take to be equal to 4. The $\epsilon$-expansion of this quantity was obtained in \cite{Diatlyk:2026eta} with a similar procedure to that described for the vectors. Alternatively, one can use that AdS$_d$ is conformally equivalent to a half-sphere and that consequently the free energy on AdS can be computed as half of the sphere free energy \cite{Sato:2021eqo,Diatlyk:2026eta}.\footnote{The same simple identification cannot be applied to scalars or vectors. The reason is that a free massless fermion in AdS effectively has a unique free boundary condition corresponding to a boundary operator of dimension $\Delta = d/2$. Vector (and scalar) fields, on the other hand, admit distinct conformal boundary conditions which result in different free energies, and cannot be captured by the single quantity $1/2 F_{S^d}$.}
%We want to compute its value at $d=4-\epsilon$ as an expansion in $\epsilon$. Generalizing the procedure applied to
%integral is divergent at large values of $u$, which makes the expansion slightly involved. To handle this, we generalize the procedure described in App.~A of Ref.~\cite{Giombi:2025pxx}, applied in that case to the free energy of conformally coupled scalars in AdS. We begin by splitting the interval of integration in two parts, [0,1) and $[1,\infty)$. The first part is convergent and can be computed numerically expanding the integrand term by term in $\epsilon$. For the second part, we replace the function multiplying the $\log (u^2)$, which we denote by $f(u)$,  with its expansion at large values of $u$, $f_\infty(u)$, in such a way that the remainder $f(u)-f_\infty(u)$ gives a finite result, and thus can be expanded in $\epsilon$ and computed numerically. The integral over $f_\infty(u)$ now captures all the divergence and can be computed analytically, by means of the integral
%\begin{equation}\label{eq:int_log}\int_1^\infty d u \ u^a \log (u^2)=\frac{2}{(1+a)^2}\,.\end{equation}
%We note that this integral would diverge for all values of $a$ in the interval $[0,\infty)$, which is why we split the original integral in two parts. Expanding the analytic result in powers of $\epsilon$ and putting all together, we thus obtain
The result for a 4-component Dirac fermion reads
\begin{equation}\label{eq:freeF}
{ F_\mathrm{free-ferm}}= \frac{11}{180 \epsilon} + 0.04943+
 0.04593\epsilon\,+\mathcal{O}\left(\epsilon^2\right)\,,
\end{equation}
which is consistent with the conformal anomaly for a fermionic  field \cite{Giombi:2014xxa} and with the fact that $F_{\mathrm{AdS}_d}-1/2 F_{S_d}$ must be finite.

\section{Free energy at the next-to-leading order}
\label{sec:free_energy_NL}
\subsection{Feynman rules in AdS}
\label{rules}
In this section, we discuss the Feynman rules in AdS$_d$ for abelian gauge theories with fermionic matter, using the notation presented in \cite{allen1986} for maximally symmetric spaces. 
In particular we denote by $\mu(x,x')$ the geodesic distance,  by $g^\nu_{\ b'}(x,x')$ the parallel propagator  transporting vectors along geodesics from $x$ to $x'$, and by $n_\nu(x,x')$ and $n_{\nu'}(x,x')$ the unit vectors  tangent to the geodesic at $x$ and $x'$ respectively:
\begin{equation}
n_{\nu}\left(x, x^{\prime}\right)=\nabla_{\nu} \mu(x, x) \quad \text { and } \quad n_{\nu^{\prime}}\left(x, x^{\prime}\right)=\nabla_{\nu^{\prime}} \mu\left(x, x^{\prime}\right).
\end{equation} 
For future convenience 
it is useful to introduce the variable $u$
\begin{equation}
u(x,x')\equiv \cosh\left({\mu(x,x')-1}\right)\,.
\label{eq:zed}
\end{equation}
which is the chordal distance between the points. It follows from symmetry arguments that the massless vector propagator can be decomposed in
\begin{equation}
G_{\nu \lambda'}(x,x')=e_0^2\left(\alpha(u)g_{\nu \lambda'}+\beta(u) n_{\nu}n_{\lambda'}\right),
\label{eq:gaugeprop}\end{equation}
where $\alpha(u)$ and $\beta(u)$ are functions of the chordal distance. Their expression, for a general $\xi$-gauge, involves a complex combination of hypergeometric functions and their derivatives. However, there exists a convenient choice of the gauge, known as the Fried-Yennie (FY) gauge, that simplifies the expression significantly \cite{Ciccone:2024guw}. In this case, by setting $\xi=d/(d-2)$ and performing some algebraic manipulations, we arrive at the following simplified form
\begin{equation}\label{eq:alphabetaD}
\begin{aligned}
    & \alpha(u)=\beta(u)=\frac{\Gamma\left(\frac{d}{2}\right)(u (2 + u)^{1 - \frac{d}{2}}}{2\pi^\frac{d}{2} ( d-3)}\,
\end{aligned}
\end{equation}
for Dirichlet BC, and
\begin{equation}
\begin{aligned}\label{eq:alphabetaN}
&\alpha(u) = \frac{ (d-3)(1+u)^2 + u(2+u) \, {}_2F_1\left( 1, \frac{3}{2}; 2 + \frac{1-d}{2}; \frac{1}{(1+u)^2} \right) }{4\pi^{\frac{d-1}{2}}(d-3)(1+u)^3 \Gamma\left( 2 + \frac{1-d}{2} \right)}\,,\\
&\beta(u) = \frac{ (3-d)u(1+u)^2 + u(2+u) \, {}_2F_1\left( 1, \frac{3}{2}; 2 + \frac{1-d}{2}; \frac{1}{(1+u)^2} \right) }{4\pi^{\frac{d-1}{2}}(d-3)(1+u)^3 \Gamma\left( 2 + \frac{1-d}{2} \right)}\,,
\end{aligned}
\end{equation}
for Neumann BC. 
The fermion propagator on  AdS$_d$ was computed in \cite{Mueck:1999efk,Giombi:2021cnr,Basu:2006ti} and reads 
\begin{equation}\label{eq:propferm}
\begin{aligned}
S^{i}_{\ j}(x,x')&=\langle{\psi}^i(x)\bar{\psi}_j(0)\rangle_\text{AdS}=-\left(\delta^i_{\ j}\ n_\mu \Gamma^\mu A_1(u)-B^i_{\ j}\  A_2(u)\right)\Lambda(x,x') \,, \\
 A_1(u)&=-\frac{\Gamma\left(\frac{d}{2}\right)u^\frac{1-d}{2}}{2^{\frac{d+1}{2}}\pi^\frac{d}{2}}\,,\quad A_2(u)=-\frac{\Gamma\left(\frac{d}{2}\right)(2 + u)^\frac{1-d}{2}}{2^{\frac{d+1}{2}}\pi^\frac{d}{2}} \,, 
   %\delta^i_{ j}\frac{\Gamma\left(\frac{d}{2}\right) \left((2 + u)^\frac{1-d}{2} \pm 
  % u^\frac{1-d}{2} n_\mu \gamma^\mu\right)}{2\pi^\frac{d}{2}}\Lambda(x,x') ,
\end{aligned}
\end{equation}
where $\Gamma^\mu$ are the gamma matrices in curved space, related to those in tangent space by the equation $\Gamma^\mu=e^\mu_a\gamma^a$, where $\{\gamma^a,\gamma^b\}=2\delta^{ab}$ and $\Lambda(x,x')$ performs the parallel transport 
\begin{equation}
\Lambda(x,x')=\frac{\gamma^a x_a \gamma_0+\gamma_0\gamma^a x'_a}{\sqrt{2 z z'(u+2)}},
\end{equation}
satisfying
    \begin{equation}
\begin{aligned}
\psi'(x')^{\alpha'}&=\Lambda(x,x')^{\alpha'}_{\ \beta}\psi(x)^{\beta}\,\\
\Lambda\left(x^{\prime}, x\right) & =\left(\Lambda\left(x, x^{\prime}\right)\right)^{-1} \\
\Gamma^{\nu^{\prime}}\left(x^{\prime}\right) & =\Lambda\left(x^{\prime}, x\right) \Gamma^\mu(x) \Lambda\left(x, x^{\prime}\right) g_{\ \mu}^{\nu^{\prime}}\left(x^{\prime}, x\right)\,.
\label{eq:parallelspin}
\end{aligned}
\end{equation}
The matrix $B$ in \eqref{eq:propferm} is a flavor matrix that squares to the identity and corresponds to the choice of boundary condition $\gamma_0 S^{i}_{\ j}(x,x')|_{z\rightarrow0}=B^i_{\ k}\ S^{k}_{\ j}(x,x')|_{z\rightarrow0}$, see \cite{Ciccone:2025dqx} for more details. In this work, we mainly restrict our attention to the maximally symmetric case $B^i_{\ k}=\mp \delta^i_{\ k}$, unless otherwise specified. %$B^i_{\ k}=\mp\delta^i_{\ k}$ and refer to \cite{Ciccone:2025dqx,DiPietro:QED3_WIP} for a discussion of more general boundary conditions.
%\footnote{More generally, in a theory with $N_f$ massless fermions, one can impose the boundary condition $\gamma_0 S^{i}_{ j}(x,x')|_{z\rightarrow0}=B^i_{k} S^{k}_{ j}(x,x')|_{z\rightarrow0}$, where $B$ is a flavor matrix that squares to the identity. In this work, we restrict our attention to the maximally symmetric case $B^i_{k}=\pm\delta^i_{k}$. We refer to \cite{Ciccone:2025dqx,DiPietro:QED3_WIP} for a discussion of more general boundary conditions.}
Substituting the explicit form of $\Lambda(x,x')$ into the propagator yields an alternative expression in terms of coordinates, which reads, in the maximally symmetric case, 
\begin{equation}
S^{i}_{\ j}(x,x')=\delta^i_{\ j}\left( \frac{\gamma^a ({x}_1-x_2)_a}{\sqrt{2 z_1 z_2}}\frac{A_1(u)}{\sqrt{u}}\mp\frac{\gamma^0\gamma^a (\bar{x}_1-x_2)_a}{\sqrt{2 z_1 z_2}}\frac{A_2(u)}{\sqrt{u+2}}-\right) \,,
\end{equation}
where $\bar{x}=(-z,\vec{x})$ is the image point with respect to the boundary.
%Note that in the massless case, Dirichlet and Neumann BC for fermionic fields become degenerate, so this expression is valid for both choices. 
Defining boundary spinors as $\widehat{\psi}(\vec{x})=z^{-\frac{d-1}{2}} \psi(\vec{x}, z \rightarrow 0)$, we have
\begin{equation}
\begin{aligned}
& S^{\partial}\ ^i _j(\vec{x},x')=\left\langle\widehat{\psi}^i\left(\vec{x}\right) \bar{\psi}_j\left(x'\right)\right\rangle=\left(\frac{-1\pm \gamma_0}{2}\right) \frac{\left(\gamma^a (x-x')_a\right) \Gamma\left(\frac{d}{2}\right)}{\sqrt{z'} \pi^{\frac{d-1}{2}}}\left(\frac{z'}{z'^2+(\vec{x}-\vec{x}')^2}\right)^{\frac{d}{2}} \\
& \bar{S}^{\partial}\ ^i _j(x,\vec{x}')=\left\langle{\psi}^i\left({x}\right) \widehat{\bar{\psi}}_j\left(\vec{x}'\right)\right\rangle=-\frac{\left(\gamma^a (x-x')_a\right) \Gamma\left(\frac{d}{2}\right)}{\sqrt{z} \pi^{\frac{d-1}{2}}}\left(\frac{z}{z^2+(\vec{x}-\vec{x}')^2}\right)^{\frac{d}{2}} \left(\frac{1\pm \gamma_0}{2}\right) 
\end{aligned}
\label{bulktoboundferm}
\end{equation}
Note that the fermion boundary condition halves the number of components, so to obtain the final boundary propagator one should project the above result onto the boundary fermion representation. 

The Feynman rules for the vertex can be extracted directly from the interacting action. We define the generic interaction vertex $\Gamma(x)$ by
\begin{equation}
S = S_\mathrm{free} - \int d^d x \sqrt{g} \, \Gamma(x)\,,
\end{equation}
such that the perturbative correction to the free energy becomes
\begin{equation}
F = F_\mathrm{free} - \sum_n\frac{1}{n!}\int d^d x_1\dots d^d x_n \sqrt{g_1}\dots{\sqrt{g_n}} \langle \Gamma(x_1)\dots  \Gamma(x_n) \rangle_{\mathrm{conn} }\,.
\end{equation}
For fermionic QED, the single fermion-photon interaction, stripping the coupling $e_0$ for convenience, reads: 
\begin{equation}
e_0\Gamma^{\mathrm{FE}}(x) = i\bar{\psi}_i\gamma^\mu\psi_i A_\mu(x)\,.
\end{equation}
\subsection{Computation of the diagram}
In the previous section, we have obtained the Feynman rules for QED with fermionic matter in AdS$_d$. We now have all the ingredients to compute the free energy at the next-to-leading order.
For $N_f$ Dirac fermions we have
\begin{equation}\label{eq:Ftotal}
\begin{split}
F=F_{\mathrm{Maxwell}}(d) +N_f F_\text{free-ferm}+F_\mathrm{curv}  +\frac{1}{2} e_0^2G_\mathrm{F}+\dots,
\end{split}
\end{equation}
where we have denoted by $G_\mathrm{F}$ the two-loop diagram
\begin{align}
    G_\mathrm{F}=\feynmandiagram[baseline=(a.base)][horizontal=a to b] {
a -- [fermion, half left] b -- [fermion,  half left] a--[gluon]b
};\, &=-\int d^d x d^d x'\sqrt{g}\sqrt{g'} \langle \Gamma^{\mathrm{FE}}(x)  \Gamma^{\mathrm{FE}}(x') \rangle_{\mathrm{conn} } \\
&=\frac{1}{e_0^2}\int d^d x d^d x'\sqrt{g}\sqrt{g'} \langle \bar{\psi_i}\gamma^\mu\psi^i A_\mu(x)  \bar{\psi_j}\gamma^\nu\psi^j A_\nu(x') \rangle _{\mathrm{conn} }\,.
\, 
\label{eq:diagram}
\end{align}
After Wick's contractions we get 
\begin{align}
&G_\mathrm{F}=- N_f  \int d^{d} x \ d^{d} x' \sqrt{g} \sqrt{g'}\ \mathrm{Tr}\left(\gamma^\mu S   \gamma
^{\mu'} S\right) \ G_{\mu\mu'}
\,,
\end{align}
which, replacing the expression for the fermion propagator in Eq.~\eqref{eq:propferm}, gives
\begin{equation}
    \begin{aligned}
        G_\mathrm{F}=- N_f \int d^{d} x \ d^{d} x' \sqrt{g} \sqrt{g'}\ &\left(A_1(u)^2n_\rho n_{\rho'}\mathrm{Tr}\left(\Gamma^{\rho'}\Lambda(x',x)\Gamma^\mu\Gamma^\rho\Lambda(x,x')\Gamma^{\mu'}\right)\right.\\ & \left.\qquad\quad+B^2A_2(u)^2\mathrm{Tr}\left(\Lambda(x',x)\Gamma^\mu\Lambda(x,x')\Gamma^{\mu'}\right)\right) G_{\mu\mu'}\,.
    \end{aligned}
\end{equation}
Since the matrix $B$ squares to the identity, this diagram evaluates to the same result regardless of the fermion boundary conditions.\footnote{Dependence on the choice of $B$ first appears in the free energy at three loops. For explicit results in $d=3$, see \cite{DiPietro:QED3_WIP}.}  Applying the rules from Eq.~\eqref{eq:parallelspin}, we move all Dirac matrices to the same point and evaluate the traces, giving
\begin{equation}\label{eq:fermdiag}
    \begin{aligned}
        G_\mathrm{F}=- N_f \mathrm{Tr}\mathbf1 \int d^{d} x \ d^{d} x' &\sqrt{g} \sqrt{g'}\ \left(\left(A_1(u)^2+A_2(u)^2\right)g^{\mu\mu'} +2A_1(u)^2n^\mu n^{\mu'})\right)G_{\mu\mu'}\,.
    \end{aligned}
\end{equation}
\paragraph{Dirichlet BC} Let us now replace the expression for the gauge propagator in Eq.~\eqref{eq:gaugeprop} and Eq.~\eqref{eq:alphabetaD} and perform contractions. As the resulting integral only depends on the geodesic distance, or equivalently on $u$, we can put $x'$ to zero and reduce the integration over $x'$ to a volume factor.
Then we convert the remaining integral in $x$ into a one-dimensional integral in the variable $u$:
\begin{equation}
\int d^{d} x \ d^{d} x' \sqrt{g} \sqrt{g'}f(u(x,x'))= \operatorname{Vol}(H^d)\ \Omega_{d-1}\int_0^\infty d u (u(u+2))^\frac{2-d}{2} f(u)\,. 
\end{equation} 
The final result is
\begin{align}
&G_\mathrm{F}^\text{D}=\ - N_f  \operatorname{Vol}(H^d)\ \Omega_{d-1}\frac{\Gamma\left(\frac{d}{2}\right)^3d}{2^d\pi^\frac{3d}{2}(d-3)}\int_0^\infty d u\ \left( u^{1-d}+(u+2)^{1-d} \right),
\end{align}
The integral of $u^{1-d}$ is  zero in dimensional regularization. To see this, one can split the interval of integration into two pieces $[0,1)$ and $[1,\infty)$ and perform different analytic continuations on them by taking $d>-2$ and $d<-2$ respectively. Summing up the two contributions, one can verify that the result vanishes. 
The remaining part of the integral is convergent for $d>2$ and can be computed analytically.
The final result is a function of $d$, which we can evaluate at $d=4$ to obtain
\begin{equation}
   G_\mathrm{F}^\text{D}=\ -N_f  \frac{1}{12\pi^2}\,.
   \label{eq:G2}
\end{equation}
Note that, contrary to what happens on the sphere, this quantity is finite. This is related to the absence of $\log e_0^2$ contributions in the leading order term of the free energy in Eq.~\eqref{eq:maxD}, or equivalently to zero mode of ghosts. 
\paragraph{Neumann BC} Let us now compute the loop diagram in Eq.~\eqref{eq:fermdiag} with Neumann BC. Replacing the expression for the gauge propagator in Eq.~\eqref{eq:gaugeprop} and Eq.~\eqref{eq:alphabetaN} and reducing the integrals into a single integral over the chordal distance, as we did for Dirichlet BC, we now get
\begin{equation}
    G_\mathrm{F}^\text{N}=\ - N_f\frac{ \cos\left(\frac{d\pi}{2}\right) \Gamma\left(\frac{d}{2}\right) \Gamma(d) \mathcal{A}(u) }{4^{d-1}\pi^{3d/2} (d-3)^2 (d-1) (1+u)^3(u(2+u))^{\frac{d}{2}+1}}\,,
    \end{equation}
    with
    \begin{equation}\begin{aligned}
    \mathcal{A} (u)&= (d-3)(1+u)^2 \left( (d-u-1)u(2+u)^d + u^d(2+u)(d+u+1) \right)\\&\ +d u (2+u) \left( u^d(2+u) + u(2+u)^d \right) \, {}_2F_1\left(1, \frac{3}{2}; \frac{5-d}{2}; \frac{1}{(1+u)^2}\right)\,.
    \end{aligned}
    \end{equation}
This integral  has divergences both for $u$ small and large. Different intervals of $d$ must be chosen to regulate them, so again we split the integral in two intervals: $u \in [0, 1)$ and $u \in [1, \infty)$. Focusing first on the interval $[0, 1)$, we decompose the integrand $g(u)$ into a small-$u$ asymptotic expansion, denoted as ${g_0}(u)$, and a remainder, $g(u)-g_0(u)$. The expansion is computed to a sufficiently high order to ensure that the integral of $g(u)-g_0(u)$ over this domain is finite.  This finite contribution is then evaluated numerically. Conversely, the integral of $g_0$ is computed via analytic continuation using  
\begin{equation}
    \int_0^1 u^a \, du = \frac{1}{1+a}\,.
\end{equation}
By expanding this analytically continued result around $d=4$, we obtain an expression which successfully isolates the pole structure. Next, we evaluate the integral over the interval $[1, \infty)$. We begin by computing the asymptotic expansion in powers of $(u+1)$ for large values of $u$, denoted as $g_\infty(u)$. As before, this expansion is carried out to a sufficiently high order such that the remainder, $g(u)-g_\infty(u)$ yields a finite integral over this domain. The integral of the expanded term $g_\infty(u)$ is then evaluated via analytic continuation, by using 
\begin{equation}
    \int_1^\infty (1+u)^a \, du = -\frac{2^{1+a}}{1+a}\,.
\end{equation}
We can expand this analytically continued result around $d=4$, now obtaining a finite result. Finally, the integral of the $g(u)-g_\infty(u)$ is computed numerically. Summing all the contributions together, we get 
\begin{equation}
   G_\mathrm{F}^\text{N}=\ N_f  \left(\frac{1}{6\pi^2\epsilon}+0.03187\right) \,.
   \label{eq:G2N}
\end{equation}
Note that this quantity is divergent, which is consistent with the presence of a $\log e_0^2$ term in the Maxwell free energy. Indeed, after replacing the bare couplings with renormalized ones, a pole arises from the logarithmic contribution that exactly cancels this divergence, as we show in the next section.
\subsection{Renormalization and free-energy at the fixed point}
The renormalization of the electric charge is fixed by the flat-space theory and reads, in minimal subtraction scheme,
\begin{equation}\label{eq:ebare}
e_0=\mu^{\frac{\epsilon}{2}}\left(e+\frac{4 N_f}{3 \epsilon} \frac{e^3}{(4 \pi)^2}+\mathcal{O}(e^5)\right).
\end{equation}
Since we are working on a curved background, we must also consider the renormalization of the curvature couplings, which are given by\ \cite{Jack:1990eb,Hathrell:1981gz}
\begin{equation}\label{eq:ren_curv}
\begin{aligned}
b_0 &= \mu^{-\epsilon}\!\left(b-\frac{11N_f+62}{360\epsilon(4\pi)^2}
+\frac{N_f}{6\epsilon}\frac{e^4}{(4\pi)^6}
+\mathcal{O}(e^6)\right),\\
c_0 &= \mu^{-\epsilon}\!\left(c+\mathcal{O}(e^6)\right).
\end{aligned}
\end{equation}
The corresponding beta functions are
\begin{equation}
\begin{aligned}
\beta_e&=-\frac{\epsilon}{2} e+\frac{4 N_f}{3} \frac{e^3}{(4 \pi)^2}+\dots,
\\
\beta_b &= \epsilon b-\frac{11N_f+62}{360(4\pi)^2}
+\frac{N_f}{2}\frac{e^4}{(4\pi)^6}
+\ldots\,,\\
\beta_c &= \epsilon c+\ldots\, ,
\end{aligned}
\end{equation}
leading to the existence of an IR-stable perturbative fixed point which, at leading order in $\epsilon$, is
\begin{equation}
    \begin{aligned}
    &e_*=\pi \sqrt{\frac{6 \epsilon}{N_f}}\,+\dots\\    &b_* = \frac{1}{\epsilon} \left( \frac{11N_f + 62}{360(4\pi)^2} - \frac{N_f}{2} \frac{e_*^4}{(4\pi)^6} + \dots \right)\,,\\
    &c_*=0+\dots,
    \end{aligned}
\end{equation}
which is the object of our study. We now replace bare couplings with renormalized ones in \eqref{eq:Ftotal}, expand up to order $e^2$, and evaluate the total free energy at the fixed point. As a non-trivial check, we see that after renormalization, the free energy is a finite function for any value of the
renormalized couplings. Specifically, the poles arising from the curvature contribution in Eq.~\eqref{eq:d3} cancel exactly against those arising from the fermion and vector one-loop determinants in Eq.~\eqref{eq:freeF} and Eq.\eqref{eq:maxD}. For Dirichlet BC, no additional poles remain. For Neumann BC, an additional pole appears in $F_\mathrm{Maxwell}$, which is exactly cancelled by the contribution in Eq.~\eqref{eq:G2N}. Computing the total free energy at the fixed point yields
\begin{equation}
\begin{aligned}
    &F_*^\mathrm{D}= \frac{11 N_f+62}{180\epsilon} -0.0134 + 0.0494 N_f + 0.2908{\epsilon}+0.0459 N_f \epsilon- \frac{9\epsilon}{64N_f}\,+\mathcal{O}\left(\epsilon^2\right),\\
   &F_*^\mathrm{N}= F_*^\mathrm{D}-1.1217 - 1.2117 \epsilon+\frac{1}{2}\log\left(\frac{N_f}{\epsilon}\right)\,+\mathcal{O}\left(\epsilon^2\right).
     %&F_*^\mathrm{N}= \frac{11 N_f+62}{180\epsilon} -0.4419 + 0.0494 N_f - 0.9209{\epsilon}+0.0459 N_f \epsilon- \frac{9\epsilon}{64N_f}+\frac{1}{2}\log\left(\frac{N_f}{\epsilon}\right)\,,
    \end{aligned}
\end{equation}
 for Dirichlet and Neumann boundary condition respectively. As expected by conformal invariance, the $\log\mu$ dependence arising after renormalization, which would become $\log(\mu R)$ if the AdS radius were taken to be different from 1, precisely cancels at the fixed point in both cases. 
%Finally, the difference between the two quantity $\delta F_*=F_*^\mathrm{N}-F_*^\mathrm{D}$ is given by
% \begin{equation}
%     \delta F_*=-0.4285 - 1.2117 \epsilon+\frac{1}{2}\log\left(\frac{N_f}{\epsilon}\right)\,.
% \end{equation}
\begin{figure} 
	\centering
\includegraphics[width=0.7\textwidth]{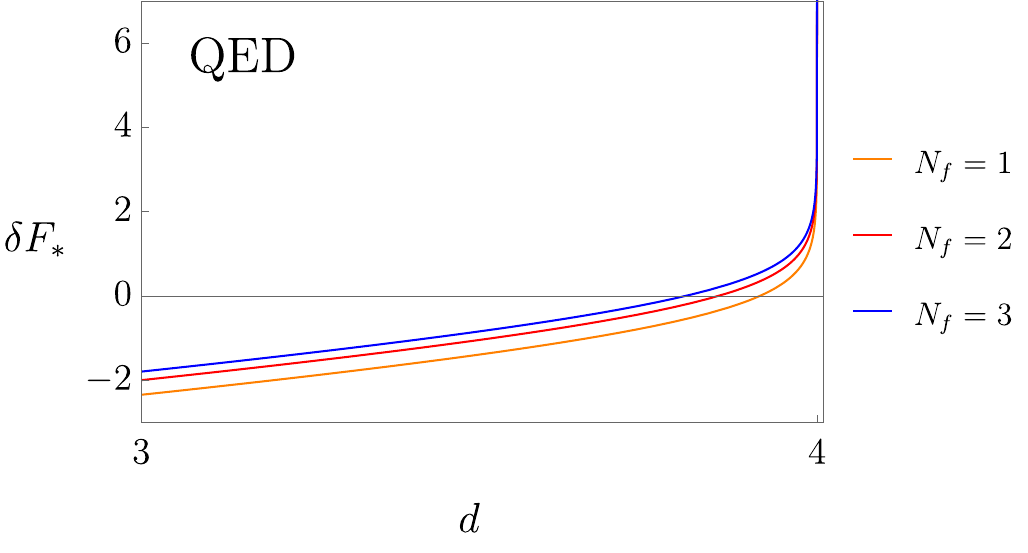} 
	\caption{Value of $\delta F_*$ at the fermionic QED fixed point in the range $3<d<4$ for different values of $N_f$. }	
	\label{fig:deltaF_Ferm}
\end{figure}
In Fig.~\ref{fig:deltaF_Ferm}, we plot $\delta \tilde{F}_*=-\sin(\pi (d-1)/2)(F_*^\mathrm{N}-F_*^\mathrm{D})$ as a function of $d$ for different values of $N_f$. We observe that in all cases $\delta F_*$ is positive and divergent close to $d=4$. This is driven by the logarithmic term arising from the ghost zero modes, which is present only in the Neumann case. As $d$ decreases, the linear term in $\epsilon$ eventually dominate, causing $\delta F_*$ to change sign at a critical dimension within the range $3 < d < 4$.

\section{Lightest boundary scalars and stability}\label{sec:BCFTferm}
Let us analyze the spectrum of the BCFT with both Dirichlet and Neumann boundary conditions. We start by considering the lightest scalar singlet operators in the free theory at $d=4$, which correspond to the displacement operators of the free Maxwell theory and the free fermionic theory when the coupling is set to zero.\footnote{We note that the fermionic bilinear $\bar{\Psi}\Psi$, which would correspond to a lighter operator, vanishes at the boundary due to the fermion boundary conditions. The operator $\bar{\Psi}\gamma_5\Psi$, where $\gamma_5$ is the chirality matrix, does not vanish at the boundary, however it is odd under parity. Here we focus on parity-even scalar operators.} Assuming Dirichlet boundary conditions for the gauge field and maximally symmetric boundary conditions for the fermions, one has
\begin{equation}
    \begin{aligned}
   & \mathcal{D}_{\rm Max}^\mathrm{D}=\frac{e^2}2 J_i J^{i}\,,\\
   & \mathcal{D}_{\rm F}^\mathrm{D}=\frac 12\mathrm{Tr}[\widehat{\bar\Psi}\overleftrightarrow{\slashed{\partial}}\widehat{\Psi}]\,,
    \label{eq:DispYM}
\end{aligned}
\end{equation}
where $\overleftrightarrow{\slashed\partial}=\gamma^i\overset{\rightarrow}{\partial}_i-\overset{\leftarrow}{\partial}_i\gamma^i$. With Neumann boundary conditions for the gauge fields, we have instead
\begin{equation}
    \begin{aligned}
    &\mathcal{D}_{\rm Max}^\mathrm{N}=-\frac1{4e^2}f_{ij}f^{ij}\,,\\
    &\mathcal{D}_{\rm F}^\mathrm{N}=\frac 12\mathrm{Tr}[\widehat{\bar\Psi}\overleftrightarrow{\slashed{D}}\widehat{\Psi}]=\frac 12\mathrm{Tr}[\widehat{\bar\Psi}\overleftrightarrow{\slashed{\partial}}\widehat{\Psi} + i\widehat{\bar\Psi}\slashed{a}\widehat{\Psi}]\,.
    \label{eq:DYMN}
\end{aligned}
\end{equation}
In both cases, the two operators have the same representation and classical dimension ($\Delta_0=4$), and therefore they mix under renormalization. At leading order, the two-point functions between the $\mathcal{D}_i$'s read
\begin{equation}
    \langle {\cal D}_i(\vec{x}_1){\cal D}_j(\vec{x}_2)\rangle=\frac{1}{(\vec{x}_{12}^{\,2})^{4}}\Big(\delta_{ij}+e^2 M_{ij}-e^2 \Gamma_{ij}\log(\vec{x}_{12}^{\,2}) + \mathcal{O}(e^4)\Big)\,, \qquad i,j=1,2\,,
\label{eq:DDExp}
\end{equation}
where we have denoted by $\mathcal{D}_{1/2}$ the operators $\mathcal{D}_{\mathrm{Max}/\mathrm{F}}$ respectively, normalized such that their two-point function at zero coupling is unit normalized. $M$ and $\Gamma$ are symmetric $2\times 2$ matrices containing information about the one-loop normalization and anomalous dimension of the operators, respectively. The matrix $\Gamma$ was computed in the non-abelian case in Ref.~\cite{Ciccone:2025dqx}. The result can be easily generalized to the abelian case by replacing the appropriate group factors. We then get
\begin{equation}
    \Gamma =
\begin{pmatrix}
0 & \mp\dfrac{N_f}{\pi^6} \\[8pt]
\mp\dfrac{N_f}{\pi^6} & \pm\dfrac{N_f}{\pi^6}
\end{pmatrix}
\end{equation}
where the signs above and below correspond to Dirichlet and Neumann BC, respectively. To compute the anomalous dimensions at $d=4$, we would simply need to diagonalize this matrix. However, because we are interested in the anomalous dimension at $d=4-\epsilon$ at the fixed point $e_*$, we must also account for the shift in the classical dimensions of the operators due to the shift in spacetime dimension, which varies depending on the chosen boundary conditions. The two-point function  in Eq.~\eqref{eq:DDExp} is therefore modified to
\begin{equation}
    \langle {\cal D}_i(\vec{x}_1){\cal D}_j(\vec{x}_2)\rangle=\frac{1}{(\vec{x}_{12}^{\,2})^{4}}\Big(\delta_{ij}+e_*^2 M_{ij}+\epsilon \ \delta\Delta_{ij}\log(\vec{x}_{12}^{\,2})-e_*^2 \Gamma_{ij}\log(\vec{x}_{12}^{\,2}) + \mathcal{O}(e^4)\Big)\,,
\end{equation}
with\begin{equation}
    \delta\Delta^\mathrm{D} =
\begin{pmatrix}
2 & 0\\
0 & 1
\end{pmatrix}\,,\quad \delta\Delta^\mathrm{N} =
\begin{pmatrix}
0 & 0\\
0 & 1
\end{pmatrix}\,.
\end{equation}
Now we simply need to replace the value of the fixed point and diagonalize the term proportional to $\log(\vec{x}_{12}^{\,2})$ to obtain the dimension of the two operators of the BCFT. Our result reads
\begin{equation}
    \begin{aligned}
    &\Delta_1^\mathrm{D}= 4 -\epsilon\,,
\\ &\Delta_2^\mathrm{D}= 4 -2\epsilon -\frac{2}{N_f}\epsilon\,,
\end{aligned}
\end{equation}
for Dirichlet BC and
\begin{equation}
    \begin{aligned}
    &\Delta_1^\mathrm{N}= 4 -\epsilon\,,\quad \quad\quad \\
    &\Delta_2^\mathrm{N}= 4 +\frac{2}{N_f}\epsilon\,,\quad \quad \quad
    \end{aligned}
\end{equation}
for Neumann BC for the gauge field. Note that generalizing to any other choice of the matrix $B$ for the fermionic boundary conditions yields the same scaling dimensions.\footnote{This can be checked by computing the mixing matrix $\Gamma$ using Eq.~(2.17) of \cite{Ciccone:2025dqx}. The only quantities potentially sensitive to the fermionic BCs are the normalization of the fermionic displacement operator $c_F$ in Eq.~(3.36) of that paper, and the related coefficient $C_{T\mathcal{D}_2}$. However, calculating $c_F$ via Wick contractions with the generic fermion propagator in \eqref{eq:propferm} shows that it is entirely independent of $B$, leaving the mixing matrix unchanged.}

Let us now comment on these results. We see that  both with Dirichlet and Neumann, we can identify the first operator with the displacement operator, which is universally present in BCFTs and possesses a protected dimension equal to $d$. Recovering the presence of this protected operator serves as a non-trivial check of our result. Let us now comment on the second operator, specifically its behavior at the physical value $\epsilon=1$. We note that with Dirichlet BC, it lies below marginality for all values of $N_f$. This one-loop result suggests that this boundary condition does not persist as a real BCFT at $d=3$, but it merges and annihilate with another boundary condition D$^*$. The situation is entirely different for Neumann BC, where the operator remains irrelevant for all values of $N_f$, suggesting that this BCFT remains real down to $d=3$.\footnote{To be more precise,  the bulk QED$_3$ theory is expected to undergo chiral symmetry breaking and lose conformality below a critical number of flavors, $N_{f,\text{crit}}$. Therefore, this conclusion physically applies only for $N_f \ge N_{f,\text{crit}}$.} This behavior is illustrated in Fig. \ref{fig:deltaF_Ferm}, where we plot the dimension of the second scalar operator for $N_f=1,2,3$ as a function of $d$. While the Neumann line remains safely above marginality in the interval $3<d<4$, the Dirichlet line crosses it for some value of $d>3$. We observe the same behavior for all positive values of $N_f$.

We report in Tab.~\ref{tab:d_critmerg} the dimension $d_\mathrm{merg}$ at which $\Delta_2^\mathrm{D}$ reaches marginality and then D would merge with D$^*$, together with the dimension $d_\mathrm{crit}$ at which $\delta\tilde{F}_* \equiv \tilde{F}_\mathrm{N} - \tilde{F}_\mathrm{D}$ changes sign, for various choices of $N_f$. We note that in all cases, $d_\mathrm{merg}$ is smaller than $d_\mathrm{crit}$. This ordering can be understood as a natural consequence of the generalized $F$-theorem in AdS discussed in the introduction. Indeed, assuming the N boundary condition represents the stable IR fixed point reached after the merger, one can argue that just before the merger occurs, there exists a relevant RG flow from D$^*$ to N. This flow is expected to be triggered by the same deformation that connects D$^*$ and D, but with the opposite sign \cite{DiPietro:2026QFTinAds}. The $F$-theorem for this flow dictates that $\tilde{F}_{\mathrm{D}^*} > \tilde{F}_{\mathrm{N}}$. Since the D and D$^*$ fixed points coincide exactly at the merger, we must have $\tilde{F}_{\mathrm{D}^*} = \tilde{F}_{\mathrm{D}}$ at $d=d_\mathrm{merg}$. Combining these facts gives $\tilde{F}_{\mathrm{D}} > \tilde{F}_{\mathrm{N}}$ precisely at the merger point. Consequently, $\delta\tilde{F}_*$ must have already changed sign prior to the merger, naturally leading to the expectation that $d_\mathrm{crit} > d_\mathrm{merg}$.

As a final remark, we note that at large $N_f$ and generic dimension $d$ with Dirichlet BC, the boundary operator $J_i$ behaves as a Generalized Free Vector with a protected dimension of $d-2$. Consequently, the composite operator $J_i J^i$ has a dimension of $2(d-2) + \mathcal{O}(1/N_f)$ and does not mix with the lightest singlet fermionic operator. Since boundary marginality is reached at $\Delta = d-1$, the operator $J_i J^i$ becomes marginal in the infinite-$N_f$ limit exactly at $d=3$, and is thus expected to trigger merger and annihilation at this dimension. This explains why $d_\mathrm{merg}$ asymptotically approaches $3$ for large $N_f$, as observed in Tab.~\ref{tab:d_critmerg}.

%As a final remark, we note that at large $N_f$ and generic dimension $d$ with Dirichlet BC, the boundary operator $J_i$ behaves as a Generalized Free Vector with a protected dimension of $d-2$. Consequently, the composite operator $J_i J^i$ has a dimension of $2(d-2) + \mathcal{O}(1/N_f)$ and does not mix with the lightest singlet fermionic operator. This is consistent with our result for $\Delta_2^\mathrm{D}$ and, since boundary marginality is reached at $\Delta = d-1$, explains why $d_\mathrm{merg}$ asymptotically approaches 3 for large $N_f$, as observed in Tab.~\ref{tab:d_critmerg}. 

\begin{table}[htpb]
    \centering
    \renewcommand{\arraystretch}{1.4}
    \begin{tabular}{|c|c|c|c|c|c|c|}
        \hline
        $N_f$ & 1 & 2 & 3 & 4 & 5 & 6 \\
        \hline
        $d_\mathrm{merg}$ & 3.667 & 3.500 & 3.400 & 3.333 & 3.286 & 3.250 \\
        \hline
        $d_\mathrm{crit}$ & 3.914 & 3.852 & 3.803 & 3.762 & 3.726 &3.696\\
        \hline
    \end{tabular}
    \caption{The dimension $d_\mathrm{merg}$ at which $\Delta_2^\mathrm{D}$ reaches marginality and the dimension $d_\mathrm{crit}$ at which $\delta\tilde{F}_*$ changes sign, evaluated for different values of $N_f$.}
    \label{tab:d_critmerg}
\end{table}

%As explained in Sec.~\ref{sec:oneloop}, there exists an RG flow from Neumann to Dirichlet BC triggered by a double-trace deformation for $d$ sufficiently close to 4, i.e., sufficiently at weak coupling so that the Dirichlet BC is still a real CFT. The generalized $F$-theorem for AdS space \FDC{Mention in the intro} then implies $\tilde{F}_{\mathrm{N}} > \tilde{F}_{\mathrm{D}}$ in this region of $d$. 

%We note that the value of $d$ at which $\Delta_2^\mathrm{D}$ reaches marginality is comparable to the value of $d$ at which $\delta\tilde{F}_*$ changes sign, for different choices of $N_f$. This may be a consequence of the monotonicity of the free energy. As explained in Sec.~\ref{sec:oneloop}, there exists an RG flow from Neumann to Dirichlet BC triggered by a double-trace deformation for $d$ sufficiently close to 4, i.e., sufficiently at weak coupling so that the Dirichlet BC is still a real CFT. The generalized $F$-theorem for AdS space \FDC{Mention in the intro} then implies $\tilde{F}_{\mathrm{N}} > \tilde{F}_{\mathrm{D}}$ in this region of $d$. Moreover, close to the value of $d$ at which D ceases to exist as a real BC because it merges with D$^*$, we expect a relevant flow from D$^*$ to N, triggered by the same deformation as that connecting D$^*$ and D, but with opposite sign \cite{DiPietro:2026QFTinAds}. This would imply $\tilde{F}_{\mathrm{D}^*} > \tilde{F}_{\mathrm{N}}$. Since D and D$^*$ coincide at the merger point, we expect $\tilde{F}_{\mathrm{D}^*} = \tilde{F}_{\mathrm{D}}|_\mathrm{merger}$, implying $\delta\tilde{F}_* = 0$ at this specific value of $d$. 
\begin{figure} 
	\centering
\includegraphics[width=0.7\textwidth]{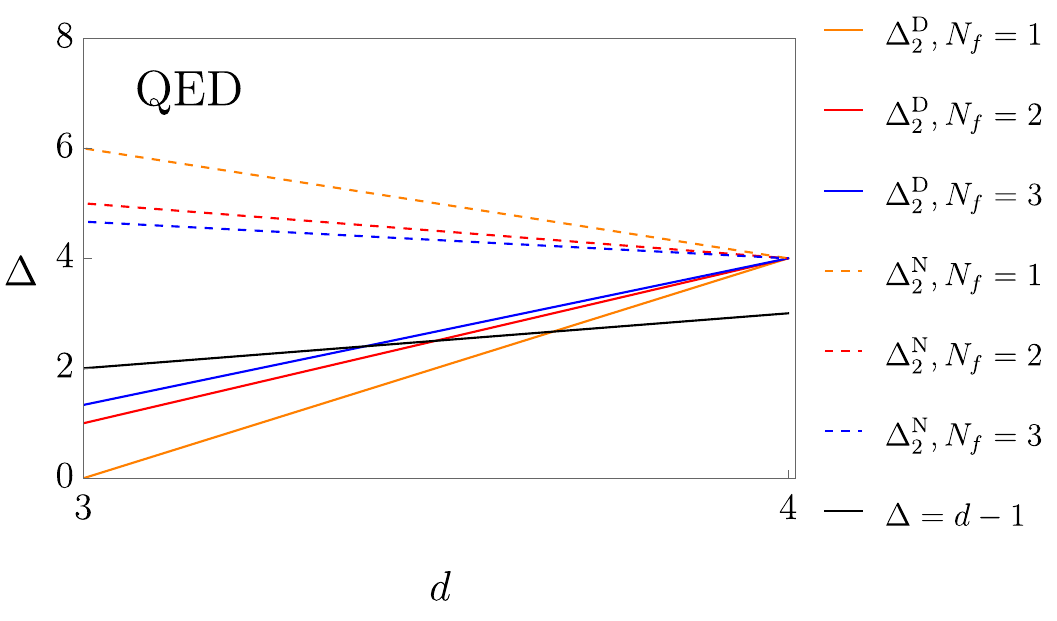} 
	\caption{Value of $\Delta_2$ for $N_f=1,2,3$ as a function of $d$ for Dirichlet (solid line) and Neumann (dashed line) BCs, compared to the value of $\Delta$ corresponding to marginality (black solid line). We see that while the red line remains safely above $\Delta=d-1$, the blue line crosses marginality for some value of $d$ between 3 and 4.  }	
	\label{fig:deltaF_Ferm}
\end{figure}

\section{Anomalous dimension with the equations of motion}
\label{sec:EOMferm}
In this section, we use the bulk equations of motion to compute the anomalous dimensions of boundary operators, following the approach of \cite{Giombi:2020rmc,Giombi:2021cnr}. Specifically, we apply this technique to compute the anomalous dimension of the boundary fermion at leading order in $\epsilon$, imposing maximally symmetric BC for the fermions and Dirichlet boundary conditions on the gauge field.\footnote{We restrict our attention to Dirichlet boundary conditions and do not repeat the computation for the Neumann case, because the boundary fermion would correspond in that case to a non-gauge-invariant operator at the boundary.} In Poincaré coordinates, the vielbein and the spin connection are given by
\begin{equation}
    e_a^\mu = z \,\delta_a^\mu, 
    \quad 
    \omega_\mu^{ab} = \frac{1}{z} \left(\delta_0^a \delta_\mu^b - \delta_0^b \delta_\mu^a \right)\,,
\end{equation}
which leads to the following expression for the Dirac operator
\begin{equation}
\gamma \cdot \nabla \Psi 
= e_a^\mu \gamma^a \left(\partial_\mu + \frac{\omega_\mu^{bc} [\gamma_b, \gamma_c]}{8} \right) \Psi
= \left(z \gamma^a \partial_a - \frac{d-1}{2} \gamma_0 \right) \Psi \,.
\end{equation}
Applying this differential operator to the bulk-to-boundary propagator of a fermion of boundary dimension $\hat{\Delta}_\Psi$, which reads
\begin{equation}
    \langle \Psi^i(x_1) \hat{\bar{\Psi}}_j(\vec{x}_2) \rangle =\delta^{i}_{\ j} B_{\Psi\hat{\Psi}} \frac{\gamma_a x_{12}^a (1 \mp \gamma_0) z_1^{\hat{\Delta}_\Psi}}{(z_1^2 + \vec{x}_{12}^2)^{\hat{\Delta}+1/2}}\,,
\end{equation}
we obtain
\begin{equation}
\gamma \cdot \nabla_1 \left\langle \Psi^i(x_1)\, \hat{\bar{\Psi}}_j(\vec{x}_2) \right\rangle
= -\left(\hat{\Delta}_\Psi - \frac{d-1}{2}\right)
\left\langle \Psi^i(x_1)\, \hat{\bar{\Psi}}_j(\vec{x}_2) \right\rangle \,.
\label{eq:eqmot1}
\end{equation}
For a free massless fermion, the equation of motion implies $\gamma \cdot \nabla \Psi = 0$, which fixes the dimension of the leading boundary spinor into $\hat{\Delta}_\Psi = (d-1)/2$. In QED, the equation of motion is modified to $
\gamma \cdot \nabla \Psi = -i e A_\mu \gamma^\mu \Psi$.
Inserting this into the right-hand side of Eq.~\eqref{eq:eqmot1}, and including an interaction vertex to make the two-point function non-vanishing, we obtain
\begin{equation}
\begin{aligned}
\gamma \cdot \nabla_1 
\left\langle \Psi^i(x_1)\, \hat{\bar{\Psi}}_j(\vec{x}_2) \right\rangle
&= e^2 \int dx \,
\left\langle A_\mu \gamma^\mu \Psi^i(x_1)\, 
\bar{\Psi}_k A_\nu \gamma^\nu \Psi^k(x)\, 
\hat{\bar{\Psi}}_j(\vec{x}_2) \right\rangle \\
&= e^2 \delta^i_{\ j} \int dx \,
G_{\mu\nu}(x_1,x)\, \gamma^\mu S(x_1,x)\, \gamma^\nu \bar{S}^\partial(x,\vec{x}_2)\,,
\end{aligned}
\label{eq:eqmot2}
\end{equation}
where in the second line we performed Wick contractions. The resulting integral contains products of several Dirac matrices, which can be rearranged using the Clifford algebra. To simplify this procedure, it is convenient to take the trace on both sides of Eq.~\eqref{eq:eqmot1} and compute the Dirac traces using the \texttt{Mathematica} package \texttt{FeynCalc} \cite{Mertig:1990an,Shtabovenko:2016sxi,Shtabovenko:2020gxv}. After this step, the right-hand side reduces to 
\begin{equation}
\left(\hat{\Delta}_\Psi - \frac{d-1}{2}\right)
\frac{2\, z_1^{\frac{d+1}{2}}}{\pi^{\frac{d}{2}} \left(\vec{x}_{12}^{\,2} + z_1^2 \right)^{\frac{d}{2}}} \,.
\end{equation}
The left-hand side is instead a linear combination of integrals of the form
\begin{equation}
I(\alpha,\beta,\gamma,\delta) = 
\int dz\, d^{d-1}\vec{x}\,
\frac{z^{\gamma}}{
\left((z-z_1)^2 + (\vec{x}-\vec{x}_1)^2\right)^{\alpha}
\left((z+z_1)^2 + (\vec{x}-\vec{x}_1)^2\right)^{\beta}
\left(z^2 + (\vec{x}-\vec{x}_2)^2\right)^{\delta}} \,.
\end{equation}
To evaluate this integral, we introduce two Feynman parameters $u_1$ and $u_2$, which make the integration over ${x}$ straightforward. Performing the integration over $u_1$ yields a hypergeometric function multiplied by a rational function. To proceed , we expand the hypergeometric function using its series definition,
\begin{equation}
{}_2F_1(a,b;c;z) = \sum_{k=0}^{\infty} 
\frac{(a)_k (b)_k}{(c)_k\, k!} z^k \,,
\end{equation}
which allows us to carry out the integration over $u_2$ term by term. The result is an integral over $z$ of the form
\begin{equation}
\begin{aligned}
I(\alpha,\beta,\gamma,\delta)
= \sum_{n=0}^\infty \int dz \;
&\frac{4^n \pi^{\frac{d-1}{2}} z^{n+\gamma} z_1^n 
\Gamma(n+\alpha) 
\Gamma\!\left(\frac{d+1}{2}+n+\gamma-\delta\right)
\Gamma\!\left(\frac{1-d}{2}+n+\alpha+\beta+\delta\right)}
{\Gamma(n+1)\Gamma(\alpha)\Gamma(n+\alpha+\beta)
\,(z+z_1)^{2(n+\alpha+\beta+\delta)+1-d}} \\
&\times {}_2\tilde{F}_1\!\left(
\delta,\,
\frac{1-d}{2}+n+\alpha+\beta+\delta;\,
\frac{d+1}{2}+n+\gamma;\,
\frac{z^2 + 2 z z_1 - (x_1 - x_2)^2}{(z+z_1)^2}
\right)\,,
\end{aligned}
\label{eq:IntThreePar}
\end{equation}
where the infinite sum originates from the series expansion of the hypergeometric function. In our case, the hypergeometric functions in Eq.~\eqref{eq:IntThreePar} reduce to polynomials, allowing the $z$ integral to be computed analytically. Summing all contributions and comparing with the right-hand side, we obtain
\begin{equation}
\hat{\Delta}_\Psi = \frac{d-1}{2} + \frac{3}{16\pi^2} e^2 \,.
\end{equation}
This result matches Eq.~(3.49) and (3.53) of \cite{Ciccone:2025dqx}, where the same quantity was computed via a direct one-loop evaluation of the boundary two-point function. At the fixed point this becomes
\begin{equation}
\hat{\Delta}_\Psi = \frac{3}{2} + \frac{9-4 N_f}{8 N_f} \epsilon \,.
\end{equation}
As a cross-check, it would be instructive to match this leading-order result with complementary computations, such as the large $N_f$ expansion of the boundary fermion dimension evaluated at $d=4-\epsilon$. 

Furthermore, it would be interesting to compute other boundary CFT data, both with Dirichlet and Neumann boundary conditions for the gauge field, either perturbatively via direct evaluations of Witten diagrams, or through non-perturbative methods. Non-perturbative approaches are particularly valuable for studying the theory with Neumann boundary conditions, as they apply to integer dimensions, specifically $d=3$, where only the interacting Neumann BCFT is expected to survive. Promising tools for this analysis include the methods mentioned in the introduction, namely Monte Carlo simulations and fuzzy sphere computations. An alternative framework, recently proposed in \cite{Loparco:2026fki,DeCesare:2026bor}, formulates the flow of QFT data in AdS as a system of ordinary differential equations. By choosing a value of $N_f$ that corresponds to a conformal bulk in the IR and tracking the evolution of the boundary operator dimensions as a function of the AdS radius, one would expect these values to plateau at large radii, directly yielding the exact BCFT dimensions.

\section{Scalar QED}
\label{sec:scalar_QED}
In this section we study QED coupled to $N_s$ conformally coupled scalars. The analysis follows the same lines of the previous sections, so we will be briefer. The action in this case is the same as Eq.\eqref{eq:b}, with $S_\mathrm{ferm}$ now replaced by
\begin{equation}
\label{eq:actionSC}
S_{\rm scal} = \int dx\,\left( D_\mu \phi_i^*  D_\mu \phi^i +
m^2 \phi_i^* \phi ^i+\lambda(\phi_i^*\phi^i)^2\right)\,,
\end{equation}
where $i=1,\dots,N_s$, $D_\mu=\partial_\mu+i A_\mu$ is the covariant derivative and the mass is such that the scalars are conformally coupled , $m^2= - d(d-2)/4$. There are two possible boundary conditions assoicated to this mass, corresponding to $\Delta_+ = d/2$, $\Delta_-=d/2-1$. We focus here on the Dirichlet case, $\Delta_+ = d/2$. 

\paragraph{Free energy at leading order}
The free energy at leading order is given by
\begin{equation}
F_\text{Free}=F_\text{Maxwell}+F_\text{scal}+F_\text{curv}\ ,
\end{equation}
where the Maxwell and curvature contributions are the same as those obtained in Sec.\ref{sec:oneloop}. The free energy for the conformally coupled scalars with Dirichlet BC can be computed from the one loop determinant in \eqref{eq:zeta} by setting $s=0$. The result was obtained as an expansion in $\epsilon$ in \cite{Giombi:2025pxx} and reads
\begin{equation}
F_\text{scal}=\frac{1}{180 \epsilon}-0.001055-0.003149 \epsilon+\mathcal{O}\left(\epsilon^2\right)\,.
\end{equation}

\paragraph{Free energy at next-to-leading order} To compute the leading order correction to the free energy, we need the scalar propagator in AdS, which for $\Delta = d/2$ reads
\begin{equation}
G(u)=\frac{\Gamma\left(\frac{d}{2}\right)\left(u^{1-\frac{d}{2}}-(u+2)^{1-\frac{d}{2}}\right)}{(2\pi)^{\frac{d}{2}}(d-2)}\,,
\end{equation}
and the interaction vertices, which can be computed from the action in \eqref{eq:actionSC},
\begin{equation}
\begin{aligned}
e_0\Gamma^{\mathrm{S1}}(x)&=-i{\phi_i}^*\overleftrightarrow\nabla_\mu\phi^i A^\mu(x) \,\\
e_0^2\Gamma^{\mathrm{S2}}(x)&={\phi^*_i}\phi^i A_\mu(x)A^\mu(x) \,.
\\
\Gamma^{\mathrm{\lambda}}(x)&=-(\phi^*_i\phi^i)^2
\end{aligned}
\end{equation}
where $\overleftrightarrow{\nabla}_\mu=\overrightarrow{\nabla}_\mu-\overleftarrow{\nabla}_\mu$.
We now have all the ingredients to compute the free energy at the next-to-leading order, which reads
\begin{equation}\label{eq:FtotalS}
\begin{split}
F=F_{\mathrm{Maxwell}}(d) +N_s F_\text{scal}+F_\mathrm{curv}  +\frac{1}{2} e_0^2G_\mathrm{S1}+e_0^2G_\mathrm{S2}+\lambda_0 G_\mathrm{\lambda}+\dots,
\end{split}
\end{equation}
where we have denoted by $G_\mathrm{S1},G_\mathrm{S2}$ and $G_\lambda$ the two-loop diagrams
\begin{align}
    &G_\mathrm{S1}=\feynmandiagram[baseline=(a.base)][horizontal=a to b] {
a -- [charged scalar, half left] b -- [charged scalar,  half left] a--[gluon]b
};\, =-\int d^d x d^d x'\sqrt{g}\sqrt{g'} \langle \Gamma^{\mathrm{S1}}(x)  \Gamma^{\mathrm{S1}}(x') \rangle_{\mathrm{conn} } 
\, \\
&G_\mathrm{S2}=\begin{tikzpicture}[baseline=(a.base)]
\begin{feynman}\vertex (a) ;
\vertex [left=0.2cm of a] ;
            \diagram* {a --  [gluon, out=45, in=-45, loop, min distance=1cm] a--  [charged scalar, out=-135, in=135, loop, min distance=1cm]a
            };
            \end{feynman}
            \end{tikzpicture}\,=-\int d^d x\sqrt{h} \langle \Gamma^{\mathrm{S2}}(x)  \rangle_{\mathrm{conn} } \\
&G_\mathrm{\lambda}=\begin{tikzpicture}[baseline=(a.base)]
\begin{feynman}\vertex (a) ;
\vertex [left=0.2cm of a] ;
            \diagram* {a --  [charged scalar, out=45, in=-45, loop, min distance=1cm] a--  [charged scalar, out=-135, in=135, loop, min distance=1cm]a
            };
            \end{feynman}
            \end{tikzpicture}\,=-\int d^d x\sqrt{h} \langle \Gamma^{\lambda}(x)  \rangle_{\mathrm{conn} } 
\label{eq:diagram}
\end{align}
After Wick's contractions we get 
\begin{align}
&G_\mathrm{S1}=
%\int d^d x d^d x'\sqrt{g}\sqrt{g'} \langle {\phi_i}^*\overleftrightarrow\nabla_\mu\phi_i A^\mu(x) (x)  {\phi_i}^*\overleftrightarrow\nabla_{\mu'}\phi_i A^{\mu'}(x) (x') \rangle _{\mathrm{conn} }\,\\&
2 N_s  \int d^{d} x \ d^{d} x' \sqrt{g} \sqrt{g'}\ (\nabla_\mu G(x,x')\nabla_\mu'G(x,x')-G(x,x')\nabla_\mu\nabla_\mu'G(x,x'))\ G^{\mu\mu'})\,,\\
&G_\mathrm{S2}=N_s  \int d^d x\sqrt{g}\ G_{\mu\mu'}(x,x)G(x,x)g^{\mu\mu'}\\
&G_\lambda=N_s(N_s+1)\int d^d x\sqrt{g}\ G(x,x)^2\,.
\end{align}
We note that the presence of double derivatives acting on the scalar propagator in $G_\mathrm{S1}$ generates contact terms which cannot be neglected \cite{DeCesare:2022obt}. To address this, one can either employ integration by parts or, as we do here,  isolate this contribution by introducing an additional contact diagram. From the equation of motion for the scalar field, we have
\begin{equation}
\nabla^2 G(x,x')+\dots=u(u+2)G''(u)+\dots=-\frac{\delta(x,x')}{\sqrt{g}}\,,
\end{equation}
where the dots do not include double derivatives and therefore do not give rise to contact terms. In our diagram double derivatives enter as
\begin{equation}
\nabla^\mu\nabla^{\mu'} G(x,x')+\dots=n^\mu n^{\mu'}u(u+2)G''(u)+\dots\ \Rightarrow\ \nabla^\mu\nabla^{\mu'} G(x,x')+\dots=-n^\mu n^{\mu'}\frac{\delta(x,x')}{\sqrt{g}}
\end{equation}
implying that the additional contribution that we need to add is
\begin{equation}
    \begin{aligned}
        &G_\mathrm{S1,c}=2N_s  \int d^d xG_{\mu\mu'}(x,x)G(x,x)n^{\mu}n^{\mu'}	\,.
    \end{aligned}
\end{equation}
After substituting the expressions for the scalar and vector propagators, we compute the integrals following the same procedure used for the fermions. We skip the details and provide directly the results. With Dirichlet boundary conditions for the gauge field, we find
\begin{align}
&G_\mathrm{S1}=-\frac{N_s}{64\pi^2}\,\\
&G_\mathrm{S1,c}=G_\mathrm{S2}=0\\
&G_\lambda=	\frac{N_s(N_s+1)}{192\pi^2}\label{eq:g2scalarD}\,,
\end{align}
We note that both $G_\mathrm{S1,c}$ and $G_\mathrm{S2}$ are equal to zero, because vector tadpoles vanish in FY gauge with Dirichlet BC. With Neumann BC, we obtain instead
\begin{align}
&G_\mathrm{S1}=N_s\left(\frac{1}{24\pi^2\epsilon}+0.01694\right)\,\\
&G_\mathrm{S1,c}=-\frac{5N_s}{48\pi^2}\\
&G_\mathrm{S2}=\frac{N_s}{24\pi^2}
\end{align}
while $G_\lambda$ remains the same as in the Dirichlet case, as it does not include vector propagators. Similar to the fermionic case, the Dirichlet diagrams are finite, whereas the Neumann case introduces a simple pole. This pole is precisely canceled after renormalization by the $\log e_0^2$ term arising from the Maxwell free energy. The renormalization of the electric and the curvature couplings  now reads \cite{Jack:1990eb,Hathrell:1981gz}
\begin{equation}\label{eq:ren_curv}
\begin{aligned}
e_0&=\mu^{\frac{\epsilon}{2}}\left(e+\frac{ N_s}{6 \epsilon} \frac{e^3}{(4 \pi)^2}+\mathcal{O}(e^5)\right).\\
\lambda_0&=\mu^{\frac{\epsilon}{2}}\left(\lambda+\frac{3 e^4 - 24 e^2 \lambda + 
 32 (N_s + 4) \lambda^2}{32 \pi^2 \epsilon}+\mathcal{O}(e^5)\right).\\
b_0 &= \mu^{-\epsilon}\!\left(b-\frac{N_s+62}{360\epsilon(4\pi)^2}
+\frac{N_s}{6\epsilon}\frac{e^4}{(4\pi)^6}
+\mathcal{O}(e^6)\right),\\
c_0 &= \mu^{-\epsilon}\!\left(c+\mathcal{O}(e^6)\right).
\end{aligned}
\end{equation}
corresponding to the beta functions
\begin{equation}
\begin{aligned}
\beta_e&=-\frac{\epsilon}{2} e+\frac{ N_s}{6} \frac{e^3}{(4 \pi)^2}+\dots,
\\
\beta_\lambda &= -\epsilon \lambda-\frac{3}{2}\frac{ e^4}{(4 \pi)^2} -12\frac{ e^2\lambda}{(4 \pi)^2}+ 16(4 + N_s) \frac{\lambda^2}{(4\pi)^2}
\\
\beta_b &= \epsilon b-\frac{N_s+62}{360(4\pi)^2}
+\frac{N_s}{2}\frac{e^4}{(4\pi)^6}
+\ldots\,,\\
\beta_c &= \epsilon c+\ldots\, ,
\end{aligned}
\end{equation}
This implies the existence of a non-trivial IR fixed point for $d<4$ and a sufficiently large number of scalars ($N_s\gtrsim 183$). 
\begin{equation}	\label{eq:fixedpointS}
    \begin{aligned}
    &e_*=\pi \sqrt{\frac{24 \epsilon}{N_s}}\,+\dots\\
    &\lambda_*=\frac{(18 + N_s + \sqrt{ N_s^2 - 180 N_s -540}) \pi^2 \epsilon}{2 N_s (4 +
N_s)}+\dots
    \\    &b_* = \frac{1}{\epsilon} \left( \frac{N_s + 62}{360(4\pi)^2} - \frac{N_s}{2} \frac{e_*^4}{(4\pi)^6} + \dots \right)\,,\\
    &c_*=\dots,
    \end{aligned}
\end{equation}
which is the object of interest of this section. 

\begin{figure} 
	\centering
\includegraphics[width=0.7\textwidth]{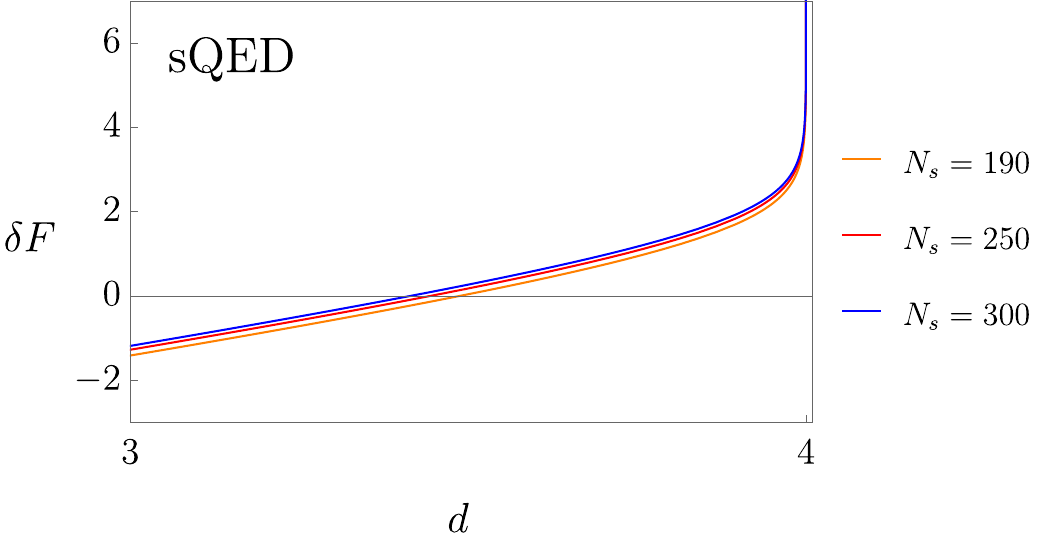} 
	\caption{Value of $\delta F_*$ at the scalar QED fixed point in the range $3<d<4$ for different values of $N_s$. }	
	\label{fig:deltaF_Scalar}
\end{figure}

\begin{figure}[h!]
	\centering
\includegraphics[width=0.7\textwidth]{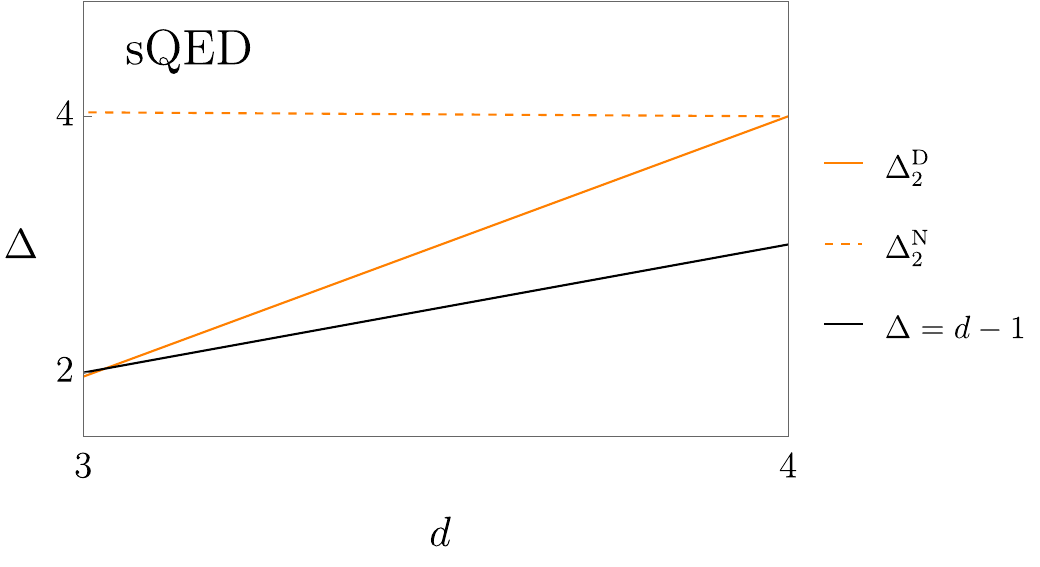} 
	\caption{Value of $\Delta_2$ for $N_s=190$ as a function of $d$ for Dirichlet (orange solid line) and Neumann (orange dashed line) BCs, compared to the value of $\Delta$ corresponding to marginality (black line). We see that while the dasehed line remains safely above $\Delta=d$, the solid line crosses marginality for some value of $d-1$ slightly above 3.}	
	\label{fig:lightestscalars_Scalar}
\end{figure}

Going back to the free energy, we can insert our results in \eqref{eq:FtotalS} and verify that all the poles cancel after replacing bare quantities with renormalized ones. We can then plug the expression of the fixed point in \eqref{eq:fixedpointS} and thus get 
\begin{equation}
\begin{aligned}
    &F_*^\mathrm{D}=  \frac{N_s + 62}{180 \epsilon} - 0.0134 - 0.0011 N_s\\&  - \frac{\epsilon( 9 + 0.7898 N_s - 0.3902 N_s^2 + 0.0005 N_s^3 - 0.0026 (N_s + N_s^2) \sqrt{N_s^2 - 180 N_s - 540} )}{N_s(N_s+4)} \\
   &F_*^\mathrm{N}= F_*^\mathrm{D}-1.8148 - 2.2117 \epsilon+\frac{1}{2}\log \left(\frac{N_s}{\epsilon}\right)\,,   \end{aligned}
\end{equation}
In Fig.~\ref{fig:deltaF_Scalar}, we plot $\delta \tilde{F}_*=-\sin(\pi (d-1)/2)(F_*^\mathrm{N}-F_*^\mathrm{D})$ as a function of the dimension $d$ for some values of $N_s$ which correspond to the existence of a real non-trivial fixed point for $\epsilon\ll 1$. We observe that, similar to the fermionic case, $\delta F_*$ is positive and divergent close to $d=4$, but it changes sign for some value between 3 and 4. 
\paragraph{Lightest boundary scalars and stability} The computation of the spectrum for the lightest scalar singlet operators in scalar QED proceeds in a completely analogous manner to the fermionic case presented in Sec.\ref{sec:BCFTferm}. Consequently, we omit the details and only highlight the relevant differences. First, the matter contribution to the mixing operators is replaced by the scalar operator $\mathcal{D}_{\rm SC} = \widehat\phi^*\widehat\phi$, for both Dirichlet and Neumann boundary conditions for the gauge field. This operator mixes with the Maxwell displacement operator $\mathcal{D}_{\rm Max}$ exactly as before. The structure of the two-point function expansion remains identical, but the one-loop mixing matrix $\Gamma$ is modified to account for the $N_s$ scalars. Also in this case we can re-adapt the results of Ref.~\cite{Ciccone:2025dqx}, but we need to add the contribution of the quartic scalar interaction, which was not taken into account in that case. The only two-point function which is affected by this interaction is $\langle\mathcal{D}_\mathrm{SC}(x_1)\mathcal{D}_\mathrm{SC}(x_2)\rangle$.
The relevant diagrams are shown in Fig.~\ref{fig:gammascalars}. Letting superscripts $(0)$ and $(1)$ denote the tree-level and one-loop contributions respectively, diagram (a) evaluates to
\begin{equation}\label{eq:ScalrPanela}
\begin{aligned}
\left.\langle \mathcal{D}_{\rm SC}(\vec{x}_1)\mathcal{D}_{\rm SC}(\vec{x}_2)\rangle^{(1)}\right|^{(a)} &= 2\langle\phi_i(\vec{x}_1) \phi^{*j}(\vec{x}_2)\rangle^{(0)}\langle\phi_j(\vec{x}_1) \phi^{*i}(\vec{x}_2)\rangle^{(1)} \\
&= -4 N_s (N_s+1) K(\vec{x}_1,\vec{x}_2)\int d^d x\sqrt{g}\, K(\vec{x}_1,x) K(x,\vec{x}_2) G(x, x)^2 \,,
\end{aligned}
\end{equation}
where the overall factor of 2 accounts for the insertion of the tadpole in the upper and lower propagator. Here, $K(\vec{x}_1,x)$ and $K(\vec{x}_1,\vec{x}_2)$ are the bulk-to-boundary and boundary-to-boundary propagators, respectively given by
\begin{equation}
K(\vec{x}_1,x) = \frac{\Gamma\left(\frac{d}{2}\right)}{\pi^\frac{d}{2}}\left(\frac{z}{z^2+(\vec{x}_1-\vec{x})^2}\right)^\frac{d}{2}\,, \quad
K(\vec{x}_1,\vec{x}_2) = \frac{\Gamma\left(\frac{d}{2}\right)}{\pi^\frac{d}{2}}\frac{1}{|\vec{x}_{12}|^d}\,.
\end{equation} 
Diagram (b) represents a contact interaction, which yields
\begin{equation}\label{eq:ScalrPanelb}
\left.\langle \mathcal{D}_{\rm SC}(\vec{x}_1)\mathcal{D}_{\rm SC}(\vec{x}_2)\rangle^{(1)}\right|^{(b)} = -2 N_s (N_s+1) \int d^d x\sqrt{g}\, K(\vec{x}_1,x)^2 K(x, \vec{x}_2)^2 \,.
\end{equation}
These diagrams can be evaluated using the procedure described in App.~D of \cite{Ciccone:2024guw}. The relevant integral evaluates to
\begin{equation}
\left. \int d^d x\sqrt{g} \left(\frac{z}{z^2+(\vec{x}_1-\vec{x})^2}\right)^\Delta \left(\frac{z}{z^2+(\vec{x}_2-\vec{x})^2}\right)^\Delta \right|_\mathrm{log} = \frac{\pi^{\frac{d-1}{2}} \Gamma \left(-\frac{d}{2}+\Delta +\frac{1}{2}\right)}{\Gamma (\Delta )|\vec{x}_{12}|^2}\,,
\end{equation}
where the subscript $\mathrm{log}$ indicates that only the terms proportional to $\log(|\vec{x}_{12}|^2)$ are retained. Substituting this result back into \eqref{eq:ScalrPanela} and \eqref{eq:ScalrPanelb}, we find that the two diagrams exactly cancel, summing to zero. 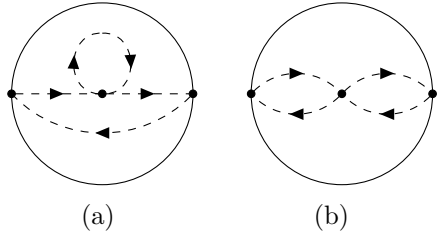
\begin{figure}[t!]
\centering\begin{subfigure}[c]{0.15\textwidth}
  \begin{minipage}{4cm}
\begin{tikzpicture}
  \begin{feynman}[inline=(a.base)]
    % Vertici principali
    \vertex (a) at (0,0) [dot] {\small{}};
    \vertex (i1) at (-1.2,0) [dot] {\small{}};
    \vertex (b) at (1.2,0) [dot] {\small{}};
    \vertex (c) at (0, 0.8); 
    \draw (0,0) circle (1.2);

    \diagram* {
      (i1) -- [charged scalar] (a),
      (a) -- [charged scalar, half left] (c) -- [charged scalar, half left] (a),
      (a) -- [charged scalar] (b)--[charged scalar, quarter left] (i1),
    };
  \end{feynman}
\end{tikzpicture}
\end{minipage}
        \caption{}
    \end{subfigure}\qquad 
        \begin{subfigure}[c]{0.14\textwidth} \begin{tikzpicture}
  \begin{feynman}[inline=(a.base)]
  \draw (0,0) circle(1.2);
    \vertex [dot](i1) at (-1.2,0){\small{ }} ;
    \vertex[dot](i2) at (0,0){\small{ }} ;
    \vertex [dot] (o2) at (1.2,0){} ;

    \diagram* {
      (i1)-- [charged scalar, quarter left] (i2)-- [charged scalar, quarter left] (o2)-- [charged scalar, quarter left] (i2) -- [charged scalar, quarter left] (i1)
    };
  \end{feynman}
\end{tikzpicture}\caption{} \end{subfigure}
        \caption{Witten diagrams that contribute to the two point function of $\mathcal{D}_{\rm SC}$  and include the quartic scalar interaction. }
        \label{fig:gammascalars}
        \end{figure} Thus, to compute $\Gamma$ we can simply take the result of \cite{Ciccone:2025dqx} and replace the group factors. This gives
\begin{equation}
\Gamma =\begin{pmatrix}
0 & \mp\dfrac{N_s}{4\pi^6} \\[8pt]
\mp\dfrac{N_s}{4\pi^6}  & \pm\dfrac{N_s}{4\pi^6} 
\end{pmatrix} 
\end{equation} 
where the upper and lower signs correspond to Dirichlet and Neumann boundary conditions, respectively. The shifts in the classical dimensions $\delta\Delta$ are again
\begin{equation}
    \delta\Delta^\mathrm{D} =
\begin{pmatrix}
2 & 0\\
0 & 1
\end{pmatrix}\,,\quad \delta\Delta^\mathrm{N} =
\begin{pmatrix}
0 & 0\\
0 & 1
\end{pmatrix}\,.
\end{equation}
The resulting anomalous dimensions are then given by 
\begin{equation}
    \begin{aligned}\Delta_1^\mathrm{D} &= 4 - \epsilon\,,\\\Delta_2^\mathrm{D} &= 4 - 2\epsilon- \frac{6}{N_s}\epsilon\,,\end{aligned}
\end{equation}
for Dirichlet BC and 
\begin{equation}
    \begin{aligned}\Delta_1^\mathrm{N} &= 4 - \epsilon\,,\\\Delta_2^\mathrm{N} &= 4 + \frac{6}{N_s}\epsilon\,,\end{aligned}
\end{equation}
for Neumann BC. As expected, in both cases we recover the exact displacement operator with protected dimension $\Delta_1 = d = 4 - \epsilon$. Furthermore, the dimension of the second operator behaves as in the fermionic theory. Under Dirichlet (Neumann) BC for the gauge field, $\Delta_2^\mathrm{D}$ lies strictly below (above) marginality at $d=3$ for all valid values of $N_s$, indicating that the BCFT is unstable (stable) in $d=3$. This behavior is illustrated in Fig.~\ref{fig:lightestscalars_Scalar}, where we plot the dimension of the second scalar operator for a representative value $N_s=190$ as a function of $d$. While the Neumann line remains above marginality in the interval $3<d<4$, the Dirichlet line crosses it for some value of $d>3$,  even if very close to $d=3$. We observe the same qualitative behavior for all values of $N_s$ that admit a real fixed point. Just as in the fermionic case, the dimension $d_\mathrm{merg}$ at which the fixed points merge is strictly smaller than the critical dimension $d_\mathrm{crit}$ at which the free energy difference between the Neumann and Dirichlet BCs changes sign. We report in Tab.~\ref{tab:d_critmergS} the values of $d_\mathrm{merg}$ and $d_\mathrm{crit}$ for selected choices of $N_s$, and note that this ordering naturally aligns with the RG flow and $F$-theorem considerations discussed above. We also note that, as in the fermionic case, at large $N_s$ and with Dirichlet BC, the composite operator $J_i J^i$ has a dimension of $2(d-2) + \mathcal{O}(1/N_s)$. This matches our result for $\Delta_2^\mathrm{D}$ and explains why $d_\mathrm{merg}$ approaches 3 for large $N_s$, as seen in Tab.~\ref{tab:d_critmergS}.

\begin{table}[htpb]
    \centering
    \renewcommand{\arraystretch}{1.4}
    \begin{tabular}{|c|c|c|c|}
        \hline
        $N_f$ & 190 & 200 & 210  \\
        \hline
        $d_\mathrm{merg}$ & 3.031 & 3.029 & 3.028 \\
        \hline
        $d_\mathrm{crit}$ & 3.484 & 3.476 & 3.469  \\
        \hline
    \end{tabular}
    \caption{The dimension $d_\mathrm{merg}$ at which $\Delta_2^\mathrm{D}$ reaches marginality and the dimension $d_\mathrm{crit}$ at which $\delta\tilde{F}_*$ changes sign, evaluated for different values of $N_s$.}
    \label{tab:d_critmergS}
\end{table}

\paragraph{Anomalous dimension with equation of motion} We now compute the leading anomalous dimension of the boundary scalar field  in the case of Dirichlet BC for the vector field, by extending the logic and computations of Sec.\ref{sec:EOMferm}. We begin by considering the bulk-to-boundary propagator \begin{equation}\langle \phi^i(x_1) \hat{\phi}_j^*(\vec{x}_2) \rangle = \delta^i_{\ j}B_{\phi\hat{\phi}} \left( \frac{z_1}{z_1^2 + (\vec{x}_1 - \vec{x}_2)^2} \right)^{\hat{\Delta}_\phi} ,
\end{equation}
where $\hat{\Delta}_\phi=d/2+\gamma$ and $\gamma$ is of order $\epsilon$ at the fixed point.  Applying the free differential operator $(\nabla^2 - m^2)$ to the bulk-to-boundary propagator, we obtain
\begin{equation}\label{eq:EOMs}(\nabla^2 - m^2) \langle \phi^i(x_1) \hat{\phi}_j^*(\vec{x}_2) \rangle = \frac{(2\hat{\Delta}_\phi - d + 1)(2\hat{\Delta}_\phi - d - 1)}{4} \langle \phi^i(x_1) \hat{\phi}_j^*(\vec{x}_2) \rangle \,,
\end{equation}
where $m^2$ is the mass of conformally coupled scalars.
On the other hand, in scalar QED the equation of motion reads
\begin{equation}(\nabla^2 - m^2) \phi^i = 2i  A^\mu \nabla_\mu \phi^i +i \nabla_\mu A^\mu \phi^i+  A_\mu A^\mu \phi^i +2\lambda \phi_i^*\phi^i \phi^i\,.\label{eq:scalar_eom_bulk}\end{equation} 
To leading order in $\epsilon$, the right-hand side of \eqref{eq:EOMs} is proportional to the anomalous dimension $\gamma_\phi$. Then, to extract the value of $\gamma$, we insert Eq.~\eqref{eq:scalar_eom_bulk} into the left-hand side and we compute its contribution at leading order in $\epsilon$. This gives
\begin{equation}\begin{aligned}(\nabla^2& - m^2) \langle \phi^i(x_1) \hat{\phi}_j^*(\vec{x}_2) \rangle = G^{\mu}_{\ \mu}(x_1, x_1) \langle \phi^i(x_1) \hat{\phi}_j^*(\vec{x}_2) \rangle+2\lambda(N_s+1)\langle \phi^i(x_1) \hat{\phi}_j^*(\vec{x}_2) \rangle\\
&+ 2 e^2 \int d^dx \sqrt{g}  G_{\mu\nu}(x_1, x) \left(\nabla_1^\mu G(x_1, x) \nabla^\nu \langle \phi^i(x) \hat{\phi}_j^*(\vec{x}_2) \rangle-\nabla_1^\mu \nabla^\nu G(x_1, x)  \langle \phi^i(x) \hat{\phi}_j^*(\vec{x}_2) \rangle\right) \\
& + e^2 \int d^dx \sqrt{g}  \nabla_1^\mu G_{\mu\nu}(x_1, x) \left( G(x_1, x) \nabla^\nu \langle \phi^i(x) \hat{\phi}_j^*(\vec{x}_2) \rangle-\nabla^\nu G(x_1, x)  \langle \phi^i(x) \hat{\phi}_j^*(\vec{x}_2) \rangle\right)\,.
\end{aligned}
\end{equation}
The evaluation of these integrals proceeds by replacing the expression for the propagators and introducing Feynman parameters to simplify the integrals over the bulk fixed point. The resulting expression is again a linear combination of the master integrals $I(\alpha, \beta, \gamma, \delta)$ discussed in Sec.\ref{sec:EOMferm}. While the values of $\alpha, \beta, \gamma, \delta$ differ due to the scalar nature of the propagators and the presence of derivatives, the analytic structure remains the same. Summing the contributions and comparing with the left-hand side, we obtain \begin{equation}
\hat{\Delta}_\phi = \frac{d}{2} +\frac{3 e^2 - 2 (N_s+1) \lambda}{16 \pi^2}\,,\end{equation}
This result is consistent with direct boundary calculations, see Eq.(4.15) of \cite{Ciccone:2025dqx} and the contribution of $\langle\phi_j(\vec{x}_1) \phi^*{}^i(\vec{x}_2)\rangle^{(1)}$ in  Eq.~\eqref{eq:ScalrPanela}, confirming the validity of the bulk EOM method.
Evaluating this result in $d=4-\epsilon$ and plugging the value of the fixed point, we get
\begin{equation}
    \hat{\Delta}_\phi =2 + \frac{\left(270 + 21 N_s - 9 N_s^2 - (1 + N_s) \sqrt{N_s^2 - 180 N_s - 540} \right) \epsilon}{16 N_s (N_s + 4)}  \,.
\end{equation}
Expanded for large value of $N_s$, this gives
\begin{equation}
    \hat{\Delta}_\phi =2 - \frac{5\epsilon}{8}  + \frac{75\epsilon}{8N_s}+ \frac{255\epsilon}{N_s^2}+\dots
\end{equation}
As in the fermionic case, it would be interesting to match this expansion against a direct large $N_s$ computation, and to extract further non-perturbative BCFT data using perturbative and non-perturbative methods.

\section*{Acknowledgments}
FDC thanks Princeton University for hospitality during the early stages of this project. We thank Pierluigi Niro for useful discussions, and Lorenzo Di Pietro for discussions and comments on the draft. We also thank the organizers of the conference ``QFT in AdS 2026'', during which this project was resumed. The research of FDC is supported by the Italian Ministry of University and Research (MUR) under the FIS grant BootBeyond (CUP: D53C24005470001) and by the
INFN “Iniziativa Specifica” ST\&FI.

\bibliography{refs}
\bibliographystyle{utphys}

\end{document}